\documentclass[twocolumn,prd,aps,superscriptaddress,showpacs]{revtex4}

\usepackage{amsmath}
\usepackage{amssymb}
\usepackage{amsfonts}
\usepackage[dvips]{color}
\usepackage{epsfig}

\def\rhom{\rho_{\psi}^*}

\def\rhocf{\rho_{\scriptsize{\alpha}}}

\def\Tcf{T_{ab}^{\text{\scriptsize{CF}}}}
\def\TKG{T_{ab}^{\text{\scriptsize{KG}}}}

\def\Tcft{T^{\text{\scriptsize{CF}}}}
\def\Sphi{S_\phi}

\def\kcut{k_\lambda}

\def\Lobs{\Lambda^{\text{\scriptsize{obs}}}}

\def\LM{\Lambda_M}

\def\LP{\Lambda_P}

\def\l{L_\lambda}
\def\L{L_\cc}
\def\lab{L_{\text{\scriptsize{lab}}}}
\def\up{L_{\text{\scriptsize{max}}}}

\def\a{\alpha}
\def\aa{\varphi}

\def\g{g_{ab}}

\def\T{T_{ab}}
\def\Tcal{T}

\def\cc{\text{\scriptsize{C}}}
\def\f{\text{\scriptsize{F}}}

\def\gb{g^{\text{\scriptsize{B}}}_{ab}}

\def\Teff{T_{ab}^{\text{\scriptsize{eff}}}}

\def\gc{\gamma_{ab}}
\def\R{R_{ab}[g]}
\def\G{G_{ab}[g]}

\def\Tc{T_{ab}}
\def\Rct{R[\gamma]}
\def\Rc{R_{ab}[\gamma]}

\def\m#1{\left\langle#1\right\rangle}
\def\mm#1{\langle#1\rangle}
\def\Teff{T_{ab}^{\text{eff}}}
\def\pd{\partial}

\def\abs#1{|#1|}

\newcommand{\kv}{\mathbf{k}}

\begin{document}

\title{Nonlinear random gravity.
I. Stochastic gravitational waves and spontaneous conformal
fluctuations due to the quantum vacuum}

\author{Charles H.-T. Wang}
\email{c.wang@abdn.ac.uk} \affiliation{SUPA Department of Physics,
University of Aberdeen, King's College, Aberdeen AB24 3UE, UK}
\affiliation{Rutherford Appleton Laboratory, STFC, Chilton, Didcot,
Oxfordshire OX11 0QX, UK}

\author{Paolo M. Bonifacio}
\email{p.bonifacio@abdn.ac.uk} \affiliation{SUPA Department of
Physics, University of Aberdeen, King's College, Aberdeen AB24 3UE,
UK}

\author{Robert Bingham}
\email{r.bingham@rl.ac.uk} \affiliation{Rutherford Appleton
Laboratory, STFC, Chilton, Didcot, Oxfordshire OX11 0QX, UK}
\affiliation{SUPA Department of Physics, University of Strathclyde,
Glasgow G4 0NG, UK}

\author{J. Tito Mendon\c{c}a}
\email{titomend@ist.utl.pt} \affiliation{CFP and CFIF, Instituto
Superior T\'{e}cnico, 1049-001 Lisboa, Portugal}
\affiliation{Rutherford Appleton Laboratory, STFC, Chilton, Didcot,
Oxfordshire OX11 0QX, UK}

\date{18 June 2008}

\begin{abstract}
We investigate the problem of metric fluctuations in the presence of
the vacuum fluctuations of matter fields and critically assess the
usual assertion that vacuum energy implies a Planckian cosmological
constant. To this end, a new stochastic classical approach to the
quantum fluctuations of spacetime is developed. The work extends
conceptually Boyer's random electrodynamics to a theory of random
gravity but has a considerably richer structure for inheriting
nonlinearity from general relativity. Attention is drawn to
subtleties in choosing boundary conditions for metric fluctuations
in relation to their dynamical consequences. We point out that those
compatible with the observed Lorentz invariance must allow for
spontaneous conformal fluctuations, in addition to stochastic
gravitational waves due to zero point gravitons. This is implemented
through an effective metric defined in terms of the random spacetime
metric modulo a fluctuating conformal factor. It satisfies an
effective Einstein equation coupled to an effective stress-energy
tensor incorporating gravitational self energy of metric
fluctuations as well as matter fields. The effective Einstein
equation is expanded perturbatively up to second order nonlinearity.
In the process of regularizing divergent integrals, a UV-cutoff is
introduced whose specific value, however, does not enter into the
resulting description of random gravity. By assuming certain
physically reasonable statistical properties of the conformal
fluctuations, it is shown that the averaged effective metric
satisfies the empty space Einstein equation with an effective
cosmological constant. This effective cosmological constant vanishes
when only the massless matter fields are included. More generally, a
finite effective cosmological constant compatible with the
observational constraints can be obtained as long as the bare masses
of the massive matter fields are nearly zero, or the conformal
invariance of matter is restored at some high energy scale.
\end{abstract}



\maketitle

\section{Introduction}



Zero point energy and vacuum fluctuations have fascinated physicists
\cite{Unruh1976, Ford1975, Ford1993, Haisch1994, Straumann1999,
Antoniadis2007} and the wider scientific audience \cite{Milonni1994,
Rugh02, Saunders2002} ever since their prediction by quantum field
theory. Although the precise quantum nature of vacuum is not fully
understood, vacuum fluctuations of the electromagnetic field have
long received support from experiments. They include the Casimir
effect \cite{Casimir1948, Casimir1948b, Mohideen1998}, Lamb shift of
atomic energy levels \cite{Welton1948, Beiersdorfer2005} and
spontaneous emission from atoms \cite{Louisell1973,
Gea-Banacloche1988, Wodkiewicz1988}. The fundamental
fluctuation-dissipation theorem \cite{Callen1951} provides a further
basis for the  vacuum fluctuations manifested as superconductive
current noise in the Josephson junctions \cite{Josephson1962,
Koch1980, Koch1982}.

Even so, elements of doubt still surround the  physical reality of
the quantum vacuum. Some argue that e.g. the Casimir effect could be
explained without the vacuum energy \cite{Jaffe2005}. Laboratory
measurements of vacuum fluctuations can at best detect differences
in the vacuum energy by modifying their boundary conditions
\cite{Doran2006}. {It} is generally believed that if vacuum energy
is real, then it must gravitate in accordance with general
relativity, and that only through the resulting gravity the net
vacuum energy could be determined. Recently, it is shown that the
part of vacuum energy responsible for the Casimir effect does indeed
gravitate \cite{Fulling2007}. However, the gravitational consequence
of the total vacuum energy remains controversial \cite{Beck2005,
Jetzer2005, Jetzer2006, Mahajan2006}.

By Lorentz invariance, {after any appropriate regularization}, the
vacuum energy can only couple to the Einstein equation via an
effective cosmological constant. There is a common perception that
an exceedingly large cosmological constant must necessarily be built
up from all ground states up to the Planck energy density scale
\cite{Weinberg1989}. While a sizable cosmological constant could
have driven the cosmic inflation in the early universe, observations
indicate its present value to be just under the critical density
value \cite{Mottola1986, Carroll2001}.

Despite the early suggested link between the vacuum energy of
elementary particles and the cosmological constant by Zel'dovich
\cite{Zeldovich1967}, a detailed mechanism is lacking. This
motivates the ``dark energy'' models such as the quintessence fields
and their extensions \cite{Weinberg1989, Carroll2001}. Leaving aside
the original mysterious cancelation of the huge vacuum energy, these
models indeed offer arguably the most popular current approach to
the ``cosmological constant problem''. Nevertheless, efforts to
account for the observed cosmic acceleration using cosmological
perturbation back-reaction without resorting to dark energy have
continued. The need for a better understanding of nonlinear metric
perturbations \cite{Kolb2006} and quantum gravity effects
\cite{Li2008} is clearly highlighted by the recent progress in this
direction.

Spacetime metric fluctuations are an integral part of the quantum
vacuum. However, a fundamental difficulty to studying these effects
is the absence of a consistent quantum theory of gravity with an
appropriate classical limit. Recently, progress has been made using
a stochastic semiclassical gravity approach
\cite{CamposVerdaguer1996, MartinVerdaguer2000,
HuRouraVerdaguer2004, Hu2004}. However, these results are limited to
linear metric fluctuations and are explicitly evaluated only for the
vacuum states of conformally invariant fields. Moffat's stochastic
gravity \cite{moffat97} is nonlinear and capable of including
general matter sources. It assumes a fluctuating gravitational
coupling strength with certain statistical properties.

In this paper, we address the issue of metric fluctuations and their
implications on cosmic acceleration without invoking additional
hypothetical fields. We shall develop a new stochastic classical
gravity approach without modifying the Einstein equation. Rather, we
seek appropriate fluctuating solutions to these equations in the
spirit of Boyer \cite{Boyer1975a}. However, we do not advocate the
stochastic classical approach to replace the ultimate need for
quantum gravity. The new approach should nonetheless allow us to
model low energy metric fluctuations due to the quantum vacuum and
assess their effect on classical spacetime at a larger scale. A
follow-up paper \cite{Bonifacio2008} will explore the possibility of
decoherence effects due to nonlinear metric fluctuations
\cite{PowerPercival2000, Wang2006}.

Classical stochastic behaviour is evident in many observed quantum
vacuum phenomena.  The effects involved may be explained using
classical electromagnetic fields satisfying Maxwell's equations, as
pointed out by Welton and others \cite{Welton1948, Cavalleri1981,
Boyer1980, IbisonHaisch1996}. The fields are however subject to
fluctuating boundary conditions, as formulated in Boyer's random
electrodynamics \cite{Boyer1975a, Boyer1975b}, to mimic vacuum
fluctuations. York championed a novel gravitational analogue in
which black hole entropy and radiance are derived from quasinormal
mode metric fluctuations \cite{York1983}. These fluctuations are
prescribed by a classical Vaidya geometry satisfying the Einstein
equation with amplitudes set to the quantum zero point level. An
improved quantum treatment of the problem using path integral is
recently provided in~\cite{York2005}.

The stochastic classical approximation to quantum fluctuations is
physically justified for low energy vacuum effects. However, new
subtleties arise when interactions and nonlinearities are included.
One hopes to derive these effects from a full quantum theory. But,
when it comes to gravity, there isn't one ready at hand! Given the
classical nature of the low energy quantum fluctuations of matter,
it seems inconsistent if they do not couple to gravity classically.
A na\"{\i}ve stochastic classical treatment of general relativity in
vacuum could be applied perturbatively as follows. For linearized
gravity, the Boyer type boundary condition results in a sea of zero
point gravitational waves. The effective stress-energy tensor of
these stochastic gravitational waves then enters into the higher
order Einstein equation \cite{Isaacson1968, Flanagan2005} as a
source. This effective energy density is positive and formally
infinite. Even after a cutoff at e.g. the Planck scale, there appear
to be problems still. The resulting effective stress-energy tensor
resembles that of a massless spin-2 radiation fluid. It cannot be
interpreted as a cosmological constant contribution, thereby
breaking the Lorentz invariance of the resulting vacuum state. Its
large energy density would lead either to a rapid expansion of the
universe as discussed by Weinberg \cite{Weinberg1989} or to a
drastic collapse of spacetime as noticed by Pauli
\cite{Straumann1999, Antoniadis2007}. It seems that adding the
vacuum energy of matter could serve only to worsen the situation.

{What has not been taken onboard in the argument above is that
nonlinear metric perturbations may themselves fluctuate. All
experiments done so far on vacuum fluctuations have clearly
demonstrated the crucial dependence of their physical consequences
on the choice of boundary conditions. A physically acceptable
boundary condition for the spacetime metric should allow for its
higher order fluctuations in the presence of nonlinear gravitational
interactions. However, cosmological observations yield constraints
on classical, averaged quantities. Hence the boundary conditions on
the nonlinear metric perturbations can practically be imposed only
after an appropriate averaging procedure. The statistical properties
of the fluctuating part of the metric are determined by requiring
that the resulting averaged classical equation matches the Einstein
equation in empty space with an effective cosmological constant.}

This paper is devoted to developing and presenting this general
framework by showing that such classical boundary conditions are
attainable if a stochastic conformal modulation of the metric
appears at the microscopic level. It is known from the canonical
analysis of general relativity \cite{DeWitt1967, Wang2005} that an
oscillatory conformal modulation of a metric induces an effective
stress-energy tensor with a negative kinetic term. {This suggests
that conformal fluctuations may yield a regularization mechanism for
the overall amount of vacuum energy.} {We will show that, provided
the conformal fluctuations satisfy certain statistical conditions,
the vacuum structure of spacetime {can be made} compatible with
Lorentz invariance at a classical scale together with an effective
cosmological constant.

In what follows, all calculations will be done locally with respect
to a physical inertial laboratory frame. When viewed at this
laboratory scale, (apparently) empty spacetime appears to be flat
and described by a Minkowski metric. We shall further assume that
there exists a cutoff scale limiting the laboratory observer access
to the microscopic structure of spacetime and its effects upon
physical probes. In particular, this cutoff scale also fixes the
energy scale which is accessible by the laboratory observer.
Throughout this paper we use the signature convention $(-,+,+,+)$
and, unless otherwise stated, {we work in dimensionless units} with
$c = G = \hbar = 1.$ {Moreover, when a tensor, e.g. $A_{ab}$, is a
functional of other tensors, e.g. $B_{cd}, C_{e}, \cdots$, we will
sometimes use the notation $A_{ab}[B, C, \cdots]$, so that the
expression $[B, C, \cdots]$ indicates functional dependence upon the
tensors $B_{cd}, C_{e}, \cdots$. When outside the brackets, the
symbol $B:=g^{ab}B_{ab}$ will denote the trace. With these
conventions, the Einstein equation with a cosmological constant
$\Lambda$ reads $G_{ab}[g] + \Lambda g_{ab} = 8\pi T_{ab}$.}


\section{Microscopic Random Gravity}

\subsection{Random scale and stochastic classical approach}

We consider the microscopic structure of spacetime at a scale $\l :=
\lambda L_P$, where $L_{P} \approx 10^{-35}$ m is the Planck scale
and where the dimensionless parameter $\lambda \gtrsim 1$. It sets
the benchmark between the full quantum gravity domain and a
semiclassical domain, in which spacetime properties still inherit
traces of the underlying quantum gravity physics, though being
expected to be treatable by semiclassical means.

To study spacetime at such a small scale we {extend} Boyer's
framework of random electrodynamics \cite{Boyer1975a,Boyer1975b} and
describe spacetime vacuum fluctuations by means of stochastic
classical fields. This kind of approach is necessary, as a definite
quantum gravity theory is still lacking. Boyer's work shows that
having a stochastic classical field results by choosing appropriate
fluctuating boundary conditions for the classical field equations.
The classical-random field can simulate quantum vacuum fluctuations
in the sense that a variety of physical phenomena, e.g. Casimir
force between two plates, can be accounted for within the
classical-random scenario.

{Even in empty spacetime the vacuum energy of matter fields is still
a source of gravity. Therefore we consider the Einstein equation for
the spacetime metric $\gc$
\begin{equation}\label{e1B}
G_{ab}[\gamma] = 8\pi \Tc[\psi,\gamma],
\end{equation}
where ${T_{ab}}$ is a model stress-energy tensor describing the
overall vacuum energy contributions coming from all matter fields,
collectively denoted by $\psi$. The metric $\gc$ and the matter
fields $\psi$ are considered to be randomly fluctuating at the scale
$\l$. Accordingly we will refer to $\l$ as to the {\emph{random
scale}} and interpret it as the typical scale above which quantum
vacuum properties can be approximately described by means of
classical stochastic fields. Below $\l$ and closer to the Planck
scale a full theory of quantum spacetime is required.

Since we are considering here vacuum metric fluctuations in an
otherwise empty universe, these can in practice be assigned as
random perturbations about some background metric $\gb$. In the
following calculations we work \emph{locally} and with respect to a
physical inertial laboratory frame, whose typical scale $\lab$ is
much larger than the random scale, yet small enough for the
background metric to appear Minkowski in the appropriate coordinate
system. With this choice we have $\gb = \eta_{ab}$ and
\begin{equation}\label{mf}
    \gamma_{ab} = \eta_{ab} + \gamma_{ab}^{(1)} + \gamma_{ab}^{(2)}
    + \ldots,
\end{equation}
where $\gamma_{ab}^{(n)}$ indicates a small fluctuating term. Here
and henceforth we follow the standard notation where a superscript
$(n)$ denotes an $n$-th order perturbation. In the subsequent
expansion of field equations, matter fields $\psi$ will be treated
as first order quantities.

The classical equation \eqref{e1B} will be analyzed explicitly in
sections \ref{LO} and \ref{QO} for the first and second order metric
perturbations. Boyer's type fluctuating boundary conditions will be
imposed on the linearized Einstein equation, so that the resulting
randomly fluctuating tensor $\gamma_{ab}^{(1)}$ can be interpreted
as describing graviton fluctuations at the random scale $\l$. Its
induced stress energy tensor is quadratic in the fluctuations and,
together with the part of $T_{ab}$ which is quadratic in the matter
fields $\psi$, it well enter the second order equation for the
higher order metric perturbations. These are not expected to
describe a physical field having its own amount of vacuum
fluctuations and, accordingly, they will not be linked to Boyer's
type random boundary conditions. They will nonetheless be
fluctuating as a result of their coupling to matter and graviton
fluctuations.}

\subsection{Macroscopic Lorentz invariance of vacuum}

Once the vacuum properties of spacetime at the random scale $\l$ are
described in terms of a stochastic classical metric, all derived
physical quantities, e.g. the Einstein tensor, are fluctuating also.
As already noted by Boyer \cite{Boyer1975a} vacuum must be Lorentz
invariant when viewed at some appropriate macroscopic scale. As a
result the only way it can possibly contribute to the Einstein
equation is {through} an effective cosmological constant term.
Lorentz invariance also implies that vacuum statistical properties
must be the same regardless of space position and direction, i.e.
\emph{homogeneity} and \emph{isotropy} hold in a statistical sense.

In order to include these properties into our formalism we want to
recover classical and smooth quantities from the fluctuating fields.
This can be obtained as a result of an averaging process. To this
end we follow the spacetime averaging procedure described in
\cite{Isaacson1968, ADM1961}, so that fluctuating tensors average to
tensors. This process involves a spacetime averaging over regions
whose typical dimensions are large in comparison to the fluctuations
typical wavelengths but smaller than the scale over which the
background geometry changes significantly. Accordingly we introduce
an averaging \emph{classical scale} $\L$ such that $\l \ll \L \ll
\lab.$ For completeness we also define a upper breakdown scale $\up
\gtrsim \lab$ as that scale at which the background deviations from
a flat geometry start to be significant.

While keeping in mind that we are here only considering `apparently'
empty spacetime, a final comment about the involved physical scales
is in order. The following hierarchy holds, with $L_P \lesssim \l
\ll \L \ll \lab \lesssim \up$. The precise characterization of the
classical scale $\L$ is that of the smallest scale at which
classical, Lorentz invariant spacetime starts to emerge as a result
of the averaging process. Though much larger than $\l$, the
classical scale $\L$ is still expected to be very small in
comparison to the laboratory scale (table \ref{scale}). The limits
of the presently suggested theory are thus clearly set: (i) below
the random scale $\l$ the random fields approximation breaks down
and a full quantum gravity theory would be needed; (ii) spacetime
starts to be smooth and classical when viewed at the classical scale
$\L$; moreover, from $\L$ and up to the upper breakdown scale $\up$,
the averaged fluctuations look like small, classical metric
perturbation upon a flat background; (iii) beyond the scale $\up$,
the approximation of flat Minkowski background breaks down.
\begin{table}[!h]
\begin{center}
\begin{tabular}{lcccc}
\hline \hline \noalign{\smallskip}  & $L_P$ & $\l$ & $\L$ & $\up$
\\
Scale & Planck & Random\hspace{0.2cm} & Classical\hspace{0.3cm} & Breakdown\\
\hline\vspace{-0.1cm} Order of & & & &
\\\vspace{-0.1cm} magn. (m)\hspace{0.2cm} & $\approx10^{-35}$ & $\gtrsim 10^{-35}$\hspace{0.2cm} &
$\gg 10^{-35}$ & $\gtrsim 1$
\\ \\\hline\vspace{-0.1cm} Physical & Quantum\hspace{0.2cm} & Random & Classical &
Classical
\\ domain & gravity & gravity &
gravity & gravity  \\
\hline Background & None & $\eta_{ab}$ & $\eta_{ab}$ &
$\gb$ \\
\hline\hline
\end{tabular}\caption{\footnotesize{\textsl{A guide to relevant physical scales.
The current formulation of the random gravity approach is supposed
to yield a valid description of spacetime from the random scale
$\l$, up to the breakdown scale $\up$. Below $\l$ a full quantum
gravity theory would be needed. Above $\up$ the flat background
approximation breaks down and one should consider the more general
case of small fluctuations upon a curved geometry
$\gb$.}}\label{scale}}
\end{center}
\end{table}

\subsection{Conformal fluctuations and the effective Einstein equation}\label{CSF}

The linearized Einstein equation with fluctuating boundary
conditions will be connected to the vacuum fluctuations of graviton
and will pose no problems. The second order equation will contain a
source term, quadratic in the fluctuating fields $\psi$ and
$\gamma_{ab}^{(1)}$, whose physical effect upon the properties of
empty spacetime must be carefully assessed. To this end, and without
loss of generality, we introduce an \emph{effective metric} $\g$,
conformally related to the metric $\gc$ via
\begin{equation}\label{e1a}
\gc = e^{2\a} \g.
\end{equation}
Hereafter we refer to $\a$ as to the \emph{conformal field}.

{The equation satisfied by the effective metric $\g$ is found in a
standard way from $G_{ab}[\gamma] = 8\pi \Tc[\psi,\gamma]$ by
re-expressing the Ricci tensor $\Rc$ and Ricci scalar $\Rct$ in
terms of $\g$, its compatible covariant derivative $\nabla_a$ and
the conformal field $\a$ as}
$$ \Rc = \R - 2\nabla_a \a_{,b} - \g\Box \a + 2 \a_{,a} \a_{,b} - 2\g \a^{,c} \a_{,c}$$
and
$$ \Rct = e^{-2\a}\left\{ R[g] -6\Box \a -6 \a^{,c} \a_{,c} \right\}. $$
Here $\a_{,a} := \nabla_a \a \equiv \pd_a \a$,
$\Box\,:=\nabla^c\nabla_c,$ while $R_{ab}[g]$ and $R[g]$ are the
Ricci tensor and Ricci scalar for the effective metric $\g.$ The
Einstein tensor $G_{ab}[\gamma] = \Rc -\frac{1}{2}\Rct \gc$ follows
as
\begin{equation}\label{e2}
G_{ab}[\gamma] = \G - 2\nabla_a \a_{,b} + 2 \g \Box\, \a + 2 \a_{,a}
\a_{,b} + \g \a^{,c} \a_{,c},
\end{equation}
and we have the following effective Einstein equation for the
effective metric $\g$
\begin{equation}\label{e3}
    \G = 8\pi(\Tc + \Sigma_{ab}).
\end{equation}
The effective stress energy tensor for the conformal field is
\begin{equation}\label{e4}
    \Sigma_{ab} := \Sigma_{ab}^{1} + \Sigma_{ab}^{2},
\end{equation}
where we have conveniently split $\Sigma_{ab}$ into the two parts:
\begin{equation}\label{s1}
\Sigma_{ab}^{1} := \frac{1}{4\pi}\left(\nabla_a \a_{,b} - \g \Box\,
\a\right),
\end{equation}
and
\begin{equation}\label{s2}
\Sigma_{ab}^{2} :=
    -\frac{1}{4\pi}\left( \a_{,a} \a_{,b} + \frac{1}{2} \g \a^{,c} \a_{,c}
    \right).
\end{equation}

The effective stress energy tensor defined above depends on $\a$
only through its derivatives. Any physical situation with $\a =
\text{const.}$ would yield $\Sigma_{ab} = 0$. In this case the two
metrics $\gamma_{ab}$ and $\g$ are equivalent and simply describe
the same spacetime with a different choice of physical units.
However, if $\a$ varies throughout spacetime, the induced non
vanishing stress energy tensor will affect the metric tensor $\g$.
In particular, conformal spacetime modulations induced by a randomly
fluctuating $\a$ would contribute, together with $\psi$ and
$\gamma_{ab}^{(1)}$, to the vacuum structure of spacetime.

Since the metric $\gc$ satisfies its own Einstein equation
$G_{ab}[\gamma] = 8\pi \Tc[\psi,\gamma]$, there is a fundamental
arbitrariness connected to the introduction of the conformal split
\eqref{e1a}. This arbitrariness has two main implications: (i) a
complete freedom in the choice of a possible dynamical equation for
$\a$, whose Boyer's type boundary conditions would lead to a
randomly fluctuating conformal field; (ii) once an equation is
chosen the statistical amplitude of the conformal fluctuations would
still be completely arbitrary.

The arbitrariness in the introduction of $\a$ will be lifted by
imposing that the \emph{averaged} second order Einstein equation for
the metric $\g$ preserves vacuum Lorentz invariance at the classical
scale and reproduces the appropriate empty space classical Einstein
equation with an effective cosmological constant. We will show
indeed that these requirements suggest a simple wave equation
holding for $\a$ and fix precisely the statistical amount of the
conformal fluctuations amplitude.}

\subsection{Characterization of the spacetime
fluctuations}\label{choiceB}

{Within our framework, spacetime at the random scale presents two
types of fluctuations: (i) \emph{conformal, scale fluctuations}
induced by the randomly fluctuating conformal field $\a$; {(ii)
\emph{effective metric fluctuations} as described by $\g$}. We now
proceed to set up the relevant stochastic properties of $\a$ and
$\g$.}

\emph{(i) Conformal fluctuations}. The field $\a$ describes small
fluctuations in the local scale of spacetime. We assign it as a
first order, stationary and stochastic field satisfying
\begin{equation}\label{as}
\m{\a}=0,\quad \quad \m{\a_{,a}}=0,
\end{equation}
where $\m{\cdot}$ indicates averaging. The particular choice
$\m{\a}=0$ corresponds to the fact that the expectation value of the
quantum field operator $\hat{\a}$ (object of the underlying quantum
theory) in the appropriate vacuum state $|0\rangle$ should vanish.

In section \ref{LO} we will interpret the first order metric
fluctuations in $\gc^{(1)}$ as describing zero point graviton
fluctuations. Accordingly, we introduce the convenient notation
\begin{equation}
    \beta_{ab}:=\gc^{(1)}.
\end{equation}
It is our choice to prescribe the conformal fluctuations in $\a$ to
be in addition to and strongly uncorrelated with the graviton
fluctuations. This choice serves to simplify the technical
derivations presented in this paper. A more detailed analysis shows
that our main results are nevertheless general enough without
assuming the statistical independence of the conformal and graviton
fluctuations. Further comments upon this particular choice are given
near the end of section \ref{LO}. Accordingly, we impose the
statistical property:
\begin{equation}\label{su0}
    \m{\a_{,\cdots} \beta_{ab,\cdots}} = 0,
\end{equation}
where ``$,\cdots$'' is a shorthand to indicate the derivatives. In
particular this implies that
\begin{equation}\label{su}
    \m{\a \beta_{ab}} = 0.
\end{equation}

When $\beta_{ab}$ is split into its independent traceless
$\beta_{ab}^*$ and trace $\beta:=\eta^{ab}\beta_{ab}$ parts as
\begin{equation}\label{td}
    \beta_{ab} = \beta_{ab}^{*} + \frac{1}{4}\eta_{ab}\beta
\end{equation}
the above relation also implies
\begin{equation}\label{suB}
    \m{\a \beta_{ab}^{*}} = \m{\a \beta} = 0.
\end{equation}
Note that indeces are here raised by $\eta_{ab}$ as we are working
with first order quantities.

\emph{(ii) Effective metric fluctuations}. Since empty spacetime
with vacuum fluctuations appears approximatively flat at the
laboratory scale $\lab$, we assume a microscopic random effective
metric of the form
\begin{equation}
\g = \eta_{ab} + g^{(1)}_{ab} + g^{(2)}_{ab}\ldots,
\end{equation}
where $\eta_{ab}$ is the Minkowski metric and $g^{(n)}_{ab}$ a
stochastic tensor describing $n$-th order fluctuations. These are
small in the sense that, in some appropriate coordinate system, they
satisfy $\abs{g^{(n)}_{ab}}\ll 1$. We shall also use the shorthand
\begin{equation}
    q_{ab}:=g^{(1)}_{ab}.
\end{equation}

For each order $n$ of the expansion, we have the classical
(non-fluctuating) part of $g^{(n)}_{ab}$ given by
\begin{equation}
g^{(n)\cc}_{ab} := \m{g^{(n)}_{ab}}
\end{equation}
and define the fluctuating part of $g^{(n)}_{ab}$ by
\begin{equation}
g^{(n)\f}_{ab} := g^{(n)}_{ab} - \m{g^{(n)}_{ab}}
\end{equation}
so that $g^{(n)}_{ab} \equiv g^{(n)\cc}_{ab}+g^{(n)\f}_{ab}$. It
follows that $g^{(n)\f}_{ab}$ has a zero mean:
\begin{equation}\label{smf}
\m{g^{(n)\f}_{ab}} = 0.
\end{equation}
The average of the random effective metric at the scale $\L$, yields
the classical effective metric
\begin{equation}\label{comp}
\m{\g} =: \g^{\cc} = \eta_{ab} + q^{\cc}_{ab} + g^{(2)\cc}_{ab} +
\mathcal{O}(3),
\end{equation}
where the terms $g^{(n)\cc}_{ab}$ could {account for} some large
scale {deviation from flat} spacetime due to vacuum fluctuations.

In terms of the metric $\gc$ we have
\begin{align*}
\gamma_{ab} &= e^{2\a}g_{ab}\\
&= \left( 1 + 2\a + 2\a^2 + \ldots\right) \times \left(\eta_{ab} +
q_{ab} + g^{(2)}_{ab} + \ldots\right).
\end{align*}
Keeping terms up to second order we have
\begin{equation}\label{mexB}
    \gamma_{ab} = \eta_{ab} +
\beta_{ab} + \gamma_{ab}^{(2)} + \mathcal{O}(3),
\end{equation}
with linear fluctuations,
\begin{equation}\label{lf}
\beta_{ab} = q_{ab} + 2\a\eta_{ab}
\end{equation}
and with nonlinear metric fluctuations given by
\begin{equation}\label{mexBB}
\gamma_{ab}^{(2)} = g^{(2)}_{ab} + 2\a^2\eta_{ab}+2{\a}q_{ab}.
\end{equation}

{After averaging, the corresponding classical metric
$\gamma_{ab}^{\cc}:=\langle\gamma_{ab}\rangle$ is, in the second
order,
\begin{align}\label{g1mB}
    \gamma_{ab}^{\cc}= \eta_{ab} + q^{\cc}_{ab} + g^{(2)\cc}_{ab} + 2\m{\a^2}\eta_{ab}
    + 2\m{{\a}q_{ab}}+ \mathcal{O}(3).
\end{align}
The mean of the cross term involving $\a$ and $q_{ab}$ gives
\begin{equation}\label{cp}
 \m{{\a}q_{ab}} = \m{{\a}\beta_{ab}} -
 2\eta_{ab}\m{\alpha^2} = -
 2\eta_{ab}\m{\alpha^2},
\end{equation}
where we have used \eqref{su}. Then
\begin{equation}\label{g1mBB}
    \gamma_{ab}^{\cc} = \left(1-2\m{\a^2}\right)\eta_{ab} + q^{\cc}_{ab} + g^{(2)\cc}_{ab} +
    \mathcal{O}(3).
\end{equation}
Comparing with \eqref{comp} we see that, apart for the re-scaling
factor $1-2\m{\a^2}$ and up to second order, the classical
properties of the spacetime metric can be described by the classical
perturbations $q^{\cc}_{ab}$ and $g^{(2)\cc}_{ab}$ of the effective
metric $\g.$} The scaling factor should be positive in order for the
metric not to become singular. Within the current second order
approximation, this implies the condition $\m{\a^2} < 1/2$. This
bound is however likely to be an artifact due to the expansion
truncation, since the full conformal factor $\exp(2\a)$ is always
positive.

{It is worth further investigating the correlation properties
between the conformal fluctuations $\a$ and the effective metric
linear perturbation $q_{ab}$. By collecting all the traceless and
trace parts in \eqref{lf} we have
\begin{equation}\label{lfII}
 \beta_{ab}^{*} + \frac{1}{4}\eta_{ab}\beta = q^{*}_{ab} +
 \frac{1}{4}\eta_{ab}(q + 8\a),
\end{equation}
where the traces are obtained as $\beta = \eta^{ab}\beta_{ab}$ and
$q= \eta^{ab}q_{ab}$. The above equation implies the two algebraic
constraints
\begin{equation}\label{ac}
    \left\{\begin{array}{l}
      q^{*}_{ab} \equiv \beta^{*}_{ab} \\
      \\
      q \equiv \beta - 8\a.
    \end{array}\right.
\end{equation}
The first constraint involving the traceless parts implies
\begin{equation}\label{tpc}
\m{\a q_{ab}^{*}} = 0,
\end{equation}
as we know from \eqref{suB} that $\a$ and $\beta^{*}_{ab}$ are
uncorrelated. From this property and equation \eqref{cp} we get
\begin{equation}\label{tpct}
    \m{\a q} = -8\m{\a^2}.
\end{equation}
The last two equations are important in that they show that the
conformal field is really just correlated to the trace part of the
linear metric perturbation $q_{ab}$.}

We conclude this section by remarking that, if the metric
perturbation $\beta_{ab}$ was put into a {\small{\texttt{TT}}}
gauge, then $\beta = 0$ identically and, as a consequence, $q =
-8\a$. In this case, apart for a multiplicative factor, $\a$ would
coincide with the trace of $q_{ab}$. It follows that, once $\beta =
0$, $q_{ab}$ cannot be put into a {\small{\texttt{TT}}} gauge as
well, or it would be $\a = 0$, leading to the trivial case $\g
\equiv \gc.$ Considering that gravitons would be described in the
{\small{\texttt{TT}}} gauge by $\beta^{*}_{ab}$ and also in view of
the identity $q^{*}_{ab} \equiv \beta^{*}_{ab}$, we can interpret
the traceless part of the metric fluctuation $q_{ab}$ as describing
graviton in the spacetime $(\mathcal{M},\g)$. We also remark that
the assignment of $\a$ as an external fluctuating field effectively
attaches physical meaning to the trace fluctuations of $q_{ab}$.

\subsection{Matter field stress-energy tensors}\label{matter}

{Vacuum carries contributions from all physical fields of nature. As
a result the random gravity framework is not complete without
inclusion of the matter fields.} These are considered to be in their
ground state. We remark that the vacuum state for the various matter
fields is well defined on a flat Minkowski background. However,
since we work at the random scale $\l$, we represent matter fields
vacuum fluctuations by means of first order and stochastic
quantities. At this stage we adopt a \emph{minimum approach} by
neglecting the non-gravitational interactions between various matter
fields components. The stress energy tensor $T_{ab}$ in \eqref{e3}
describes matter fields and, ideally, carries contributions from all
sectors of the Standard Model. Then $T_{ab} = \sum_{j}{T}^{j}_{ab}$,
where the index $j$ runs over \emph{all} matter fields.

The detailed microscopic expression of the generic component
${T}^{j}_{ab}$ will depend \emph{quadratically} upon the
corresponding fluctuating matter field, as well as on the random
metric $\gc = \eta_{ab} + \sum_n \gc^{(n)}$. As a result the stress
energy tensor at the random scale $\l$ is a also a stochastic
quantity. The dependence on the $\gc^{(n)}$ would account for the
back-reaction of gravity fluctuations upon matter fields. However,
as long as we work up to second order, only the flat classic
background $\eta_{ab}$ will appear but \emph{not} the metric
fluctuations $\gc^{(n)}$. In this sense it is like having
fluctuating fields on a Minkowski background and the effect of
gravity upon the stress energy tensor would only appear as a third
order effect. Within this framework, in an appropriate coordinate
system, matter fields stress energy tensor component are quadratic
fluctuating quantities defined on a flat background.

The corresponding energy density contribution can be defined at the
classical scale $\L$ in a statistical sense through an averaging
procedure. Provided the high frequency components are cutoff at the
random scale $\l = \lambda L_P$, the average will be well defined
and finite. Then the quantity $\m{T_{ab}}$ gives rise to a
macroscopic stress-energy tensor at the classical scale $\L$.
Because of homogeneity and isotropy holding at the classical scale,
its most general form is
\begin{equation}\label{set}
    \m{T_{ab}} = \left(
               \begin{array}{cccc}
                 \rho & 0 & 0 & 0 \\
                 0 & p & 0 & 0 \\
                 0& 0 & p & 0 \\
                 0 & 0 & 0& p \\
               \end{array}
             \right)
\end{equation}
where $\rho := \m{{T}_{00}}$ is the energy density and $ p :=
\frac13 \m{{T}^i{}_i} $, $i=1,2,3$ is the pressure. The dominant
energy condition, i.e. $\rho \geq 0$ and $\rho \geq p$, is normally
thought to be valid for all known reasonable forms of matter
\cite{Hawking&Ellis1973}, at least as long as the adiabatic speed of
sound $dp / d\rho$ is less than the speed of light. This is true for
massless fields since, in this case, $\rho = 3p$. In appendix
\ref{A1} we show that the stronger condition $\rho > 3p$ is
satisfied by massive fields in their vacuum state, at least in the
ideal case in which interactions can be neglected and the field
masses are much smaller than the Planck mass.

\subsection{{From random gravity to classical
gravity}}

The physical picture is such that, at the random scale $\l = \lambda
L_P$, spacetime and matter fields are randomly fluctuating. As a
result we have the \emph{random Einstein equation} for the effective
metric $\g$
\begin{equation}\label{mree1}
\G = 8\pi \Teff[\psi,\a,g],
\end{equation}
with $\psi$ denoting collectively all matter fields and where the
effective stress energy tensor induced by the conformal fluctuations
is
\begin{equation}\label{eff}
\Teff[\psi,\a,g] = \Sigma_{ab}^{1}[\a,g] + \Sigma_{ab}^{2}[\a,g] +
\Tc[\psi,\alpha,g],
\end{equation}
the linear and quadratic parts in $\a$ being given in equations
\eqref{s1} and \eqref{s2}.

At the classical scale $\L$ and above, the averaged microscopic
equation will yield a corresponding classical equation in terms of
smooth, non-random quantities $ \m{\G} = 8\pi
\m{{\Teff[\psi,\a,g]}}. $ We will now proceed to apply an order by
order expansion scheme to equation \eqref{mree1} and study under
which conditions it takes the only form which is compatible with
Lorentz invariance of vacuum, i.e. that of the empty space classical
Einstein equation with an effective cosmological constant term.

\section{First order analysis of the random Einstein
equation}\label{LO}

\subsection{Linear solution and graviton vacuum fluctuations}

We now proceed to linearize the microscopic random Einstein equation
\eqref{mree1} and study the structure of its effective stress energy
tensor, including vacuum conformal fluctuations. The terms
$\Sigma_{ab}^{2}$ and $\Tc$ contain fluctuations starting from
second order so that only the term $\Sigma_{ab}^{1}$ contributes.
Using equation \eqref{s1} we have the linearized random Einstein
equation
\begin{equation}\label{1storder}
    G^{(1)}_{ab}[q] = 2\pd_a\pd_b\a -
    2\eta_{ab}\partial^c\partial_c
    \a,
\end{equation}
where $G^{(1)}_{ab}$ is the usual linear operator resulting from the
linearized Einstein tensor \cite{Flanagan2005} and $q_{ab}$
represents the first order metric fluctuations. Here the indices are
shifted using $\eta_{ab}$ and $\eta^{ab}$.

This equation can be simplified. Indeed, from $q_{ab} = \beta_{ab} -
2\a\eta_{ab}$, it is readily verified using the explicit expression
for $G^{(1)}_{ab}$ given in appendix \ref{A2} that
\begin{equation}\label{yyy}
G^{(1)}_{ab}[q] \equiv G^{(1)}_{ab}[\beta - 2\a\eta] =
G^{(1)}_{ab}[\beta] + 2\pd_a\pd_b\a -
    2\eta_{ab}\partial^c\partial_c
    \a.
\end{equation}
Comparing with \eqref{1storder} we see that the first order
approximation to the microscopic Einstein equation takes the form
\begin{equation}\label{foee}
    G^{(1)}_{ab}[\beta] = 0,
\end{equation}
with $\beta_{ab} = q_{ab} + 2\a\eta_{ab}$ and
\begin{equation}\label{mean}
\m{\beta_{ab}} = \m{q_{ab}} \equiv q^{\cc}_{ab}.
\end{equation}

In the Lorentz gauge, equation \eqref{foee} is the ordinary wave
equation and it is usually considered to describe weak gravitational
waves (GWs). Within our current random framework, it can be assigned
fluctuating boundary conditions. As a result, the fluctuating tensor
$\beta_{ab}$ is thought to represent the vacuum fluctuations of
\emph{graviton}. Note that, as explained in section \ref{CSF}, the
conformal field $\a$ can in principle be assigned arbitrarily. Once
this is done, $q_{ab}$ is fixed from $q_{ab} = \beta_{ab} -
2\a\eta_{ab}$ in such a way that \eqref{1storder} is automatically
satisfied. We remark how the conformal fluctuations act in equation
\eqref{1storder} as a fluctuating forcing source term, implying that
the first order effective metric fluctuations $q_{ab}$ correlate to
$\a$. This observation rules out the possibility $\m{\a q_{ab}}=0$
and motivates our choice $\m{\a\beta_{ab}}=0$ in section
\ref{choiceB} as, in fact, a physically reasonable scenario.

The corresponding equation holding at the classical scale $\L$ is
found by taking the mean in \eqref{1storder} or \eqref{foee}. Since
$G^{(1)}_{ab}$ is a linear operator it commutes with the average
operation \cite{Flanagan2005} and $ \langle
G^{(1)}_{ab}[\beta]\rangle = G^{(1)}_{ab}[\langle\beta\rangle],$
implying
\begin{equation}
\quad G^{(1)}_{ab}[q^{\cc}] = 0,
\end{equation}
where we have used \eqref{mean}. The first order, classical, metric
correction $q^{\cc}_{ab}$ would describe classical gravitational
waves propagating on a Minkowski background. Since we want to
characterize spacetime vacuum properties only, we set these GWs to
zero by choosing
\begin{equation}\label{1st}
q^{\cc}_{ab} = 0.
\end{equation}
This represents the first order solution to our problem. We have
here the important result that, in the first order, vacuum
fluctuations do not give any visible effect on the structure of
spacetime.

\subsection{Drift of classical spacetime due to non linear vacuum
effects}

After taking into account the first order solution, the metric
structure follows from equations \eqref{mexB}-\eqref{mexBB} as
\begin{equation}\label{mex}
    \gamma_{ab} = \eta_{ab} +
\beta_{ab}^{\f} + \gamma_{ab}^{(2)} + \mathcal{O}(3),
\end{equation}
with the zero mean, linear fluctuations,
\begin{equation}\label{lbl}
\beta_{ab}^{\f} = q^{\f}_{ab} + 2\a\eta_{ab}
\end{equation}
describing graviton vacuum fluctuations, and with second order
fluctuations
\begin{equation}
\gamma_{ab}^{(2)} = g^{(2)}_{ab} + 2\a^2\eta_{ab}+2{\a}q^{\f}_{ab}.
\end{equation}
After averaging, the full classical metric follows as
\begin{equation}\label{g1m}
    \m{\gamma_{ab}} = \left(1 - 2\m{\a^2}\right)\eta_{ab} +
    g^{(2)\cc}_{ab} +
    \mathcal{O}(3),
\end{equation}
where we have used \eqref{cp}. This result shows that a deviation
from a flat Minkowski background can arise as a second order drift
effect due to nonlinear vacuum fluctuations. The classical equation
satisfied by $g^{(2)\cc}_{ab}$ is found by a second order analysis
of the averaged random Einstein equation.

\section{Second order analysis of the random Einstein
equation}\label{QO}

\subsection{Second order equation and gravitons effective stress energy tensor in vacuum}

Expanding the fluctuating Einstein tensor in the random equation
\eqref{mree1} up to second order yields the result
\cite{Flanagan2005}
\begin{equation}\label{eq}
G_{ab} = G^{(1)}_{ab}[q] + G^{(1)}_{ab}[g^{(2)}] + G^{(2)}_{ab}[q] +
\mathcal{O}(3),
\end{equation}
where $G^{(1)}_{ab}$ is the linear operator introduced above while
$G^{(2)}_{ab}$ is quadratic in its argument \cite{MTW} and its
explicit structure is reported in appendix \ref{A2}.

Because observations of the spacetime metric involve classical,
macroscopic scales, we want to analyze the structure of the averaged
second order equation. Taking the mean in \eqref{eq}, the first term
vanishes by virtue of the first order solution. Then, using the fact
that $q_{ab}=q^{\f}_{ab}$ we get
\begin{equation}\label{lhs}
\m{G_{ab}} = G^{(1)}_{ab}[g^{(2)\cc}] + \m{G^{(2)}_{ab}[q^{\f}]} +
\mathcal{O}(3).
\end{equation}
Using $q_{ab}^{\f} = \beta_{ab}^{\f} - 2\a\eta_{ab}$ we obtain
\begin{equation}\label{lhs1}
\m{G_{ab}} = G^{(1)}_{ab}[g^{(2)\cc}] +
\m{G^{(2)}_{ab}[\beta^{\f}]}+ \m{G^{(2)}_{ab}[2\alpha\eta]}.
\end{equation}
A fourth term deriving from $\langle G^{(2)}_{ab}[q^{\f}]\rangle$
and containing products of the derivatives of $\beta_{ab}^{\f}$ and
$\a$ averages to zero because of the fact that these two quantities
are strongly uncorrelated as expressed in \eqref{su0}.

The second term in \eqref{lhs1} gives rise to the effective
stress-energy tensor
\begin{equation}\label{tgw}
\m{T_{ab}^{\text{GW}}[\beta^{\f}]} :=
-\frac{1}{8\pi}\m{G^{(2)}_{ab}[\beta^{\f}]},
\end{equation}
which is quadratic in the first order fluctuations $\beta_{ab}^{\f}$
and acts, together with matter and conformal field fluctuations
effective tensors, as an extra source for the classical, second
order metric drift term $g^{(2)\cc}_{ab}$.

The structure of this effective stress-energy tensor has been
studied extensively in \cite{Isaacson1968}, where it is shown to be
positive definite, traceless, and well defined in describing the
energy content of linear gravitational fluctuations, as long as the
wavelengths of the fluctuations involved are shorter than the
typical scale over which the geometry of the background metric
varies significantly. In our case this means that the graviton
fluctuations wavelength contained in $\beta_{ab}^{\f}$ should be
shorter than the breakdown scale $\up$. This would technically imply
an infrared cutoff. However, since in the current version of the
theory we are working with a flat background we will ignore this.
Within our present framework, $\m{T_{ab}^{\text{GW}}}$ will be
connected to the energy density due to graviton vacuum fluctuations.
Given the properties of $\m{T_{ab}^{\text{GW}}}$ and the fact that
it effectively describes gravitons as massless spin-2 particles, we
can avoid considering its explicit averaged structure by simply
including gravitons within the collection of matter fields, as
described collectively by $\psi$.

With this in mind, the averaged second order equation takes the form
\begin{eqnarray}\label{2nd}
G^{(1)}_{ab}[g^{(2)\cc}] = 8\pi\left\langle
\Tcf{}^{(2)}[\a,\beta^{\f}] + \Tc^{(2)}[\psi]\right\rangle
\end{eqnarray}
where $\psi$ now includes gravitons as represented by
$\beta_{ab}^{\f}$ and the superscripts $^{(2)}$ indicate second
order quantities. The first term on the r.h.s. is the second order,
averaged effective stress-energy tensor due to the conformal
fluctuations, given by
\begin{align}\label{mtcf}
\m{\Tcf{}^{(2)}[\a,\beta^{\f}]}:=&\m{\Sigma_{ab}^{1(2)}[\a,\beta^{\f}]\nonumber
+ \Sigma_{ab}^{2(2)}[\a]}\\
&-\m{\frac{1}{2\pi}G^{(2)}_{ab}[\a\eta]}.
\end{align}

\subsection{Effective stress energy tensor analysis}

\subsubsection{Conformal field}

We analyze the three contributions to $\m{\Tcf{}^{(2)}}$ given in
\eqref{mtcf}. More details on the derivations of the following
equations are given in appendix \ref{A2}. From \eqref{s1} the first
term can be seen to contain second order terms involving products of
$\a$ and $q_{ab}^{\f}$. Using equation \eqref{lbl} to replace
$q_{ab}^{\f}$ and using the fact that $\a$ and $\beta_{ab}^{\f}$ are
uncorrelated, this term turns out to be independent of
$\beta^{\f}_{ab}$ and takes the form:
\begin{equation}\label{S12}
\m{\Sigma_{ab}^{1(2)}[\a,\beta^{\f}]}=\frac{1}{4\pi}\m{
2\a_{,a}\a_{,b} +\eta_{ab}\a^{,c}\a_{,c}}.
\end{equation}
The second term in \eqref{mtcf} follows from \eqref{s2} replacing
$\g$ with $\eta_{ab}$:
\begin{equation}\label{S22}
\m{\Sigma_{ab}^{2(2)}[\a]}=-\frac{1}{4\pi}\m{ \a_{,a} \a_{,b} +
\frac{1}{2} \eta_{ab} \a^{,c} \a_{,c}}.
\end{equation}
Using the explicit form of the non linear operator $G^{(2)}_{ab}$,
the third term can be calculated to be:
\begin{equation}\label{qqq}
 -\frac{1}{2\pi}\m{G^{(2)}_{ab}[\a\eta]} =
 -\frac{3}{4\pi}\m{\a_{,a}\a_{,b}}.
\end{equation}
Collecting the above results, equation \eqref{mtcf} becomes
\begin{equation}\label{tr}
\m{\Tcf{}^{(2)}[\alpha]} = -\frac{1}{2\pi}\m{ \a_{,a}\a_{,b} -
\frac{1}{4}\eta_{ab}\a^{,c}\a_{,c} }.
\end{equation}

Considering the stochastic properties of the conformal fluctuations
in different spacetime directions to be uncorrelated and since
$\m{\a_{,c}}=0$ we have $ \m{\a_{,a} \a_{,b}} =
\m{\a_{,a}}\m{\a_{,b}} = \delta_{ab}\m{\a_{,a}^2}, $ where
$\delta_{ab}$ is the four dimensional Kronecker symbol. Isotropy at
the classical scale implies that, for all three space directions,
\begin{equation}\label{is}
    \m{\a_{,i}^2} = \m{\abs{\nabla\a}^2}/3,\quad i=1,2,3,
\end{equation}
where $\m{\abs{\nabla\a}^2} = \m{\a_1^2} + \m{\a_2^2} + \m{\a_3^2}.$
Then it is readily verified that the average of the tensor in
\eqref{tr} has the structure
\begin{equation}\label{tst}
    \m{\Tcf{}^{(2)}[\a]} = \left(
      \begin{array}{cccc}
        \rhocf & 0 & 0 & 0 \\
        0 & \frac{\rhocf}{3} & 0 & 0 \\
        0 & 0 & \frac{\rhocf}{3} & 0 \\
        0 & 0 & 0 & \frac{\rhocf}{3} \\
      \end{array}
    \right),
\end{equation}
where
\begin{align}\label{ccc1}
  &\rhocf := -\frac{3}{8\pi}\left(\m{\a_{,0}^{\,2}} +
\frac{\m{\abs{\nabla\a}^2}}{3}\right) \leq 0
\end{align}
Quite remarkably it is traceless:
\begin{align}\label{ccc2}
\m{\Tcft{}^{(2)}[\a]} = 0.
\end{align}
The classical, average quantity $\rhocf$  corresponds to an
effective, negative energy density connected to the vacuum conformal
fluctuations.

\subsubsection{Matter fields}

The second order, averaged matter stress-energy tensor is given in
equation \eqref{set} in terms of the matter vacuum energy density
$\rho$ and pressure $p$. As already discussed, it now also includes
gravitons as massless particles. Performing a trace decomposition it
has the structure
\begin{align}\label{tst2}
    \m{\T^{(2)}[\psi]} =& \left(
      \begin{array}{cccc}
        \rhom & 0 & 0 & 0 \\
        0 & \frac{\rhom}{3} & 0 & 0 \\
        0 & 0 & \frac{\rhom}{3} & 0 \\
        0 & 0 & 0 & \frac{\rhom}{3} \\
      \end{array}
    \right)
+
\frac{1}{4}\eta_{ab}\m{\Tcal^{(2)}}
\end{align}
where
\begin{align}
  &\rhom := \frac{3}{4}\left(\rho + p\right),\label{aaa1}
\end{align}
is the energy density of the traceless part and
\begin{align}
  &\m{\Tcal^{(2)}} = 3p - \rho\label{aaa2}.
\end{align}
is the trace part, with $\Tcal^{(2)} = \eta^{ab}\T^{(2)}$.

To proceed, it is conceptually useful to separate the fields $\psi$
into massless fields (including gravitons) and massive fields,
collectively denoted by $\Psi$ and $\Upsilon$ respectively. Massless
fields contained in $\Psi$ must have a traceless stress-energy
tensor. This implies the usual equation of state
\begin{equation}\label{eqsms}
\rho_{\Psi} = 3p_{\Psi},
\end{equation}
where $\rho_{\Psi}$ and $p_{\Psi}$ denote the contribution to the
vacuum energy density and pressure coming form all massless fields,
including graviton. In a similar way $\rho_{\Upsilon}$ and
$p_{\Upsilon}$ will indicate the corresponding contributions from
the massive fields. With this notation equations \eqref{aaa1} and
\eqref{aaa2} can be re-expressed as
\begin{align}\label{bbb1}
  &\rhom := \rho_{\Psi} + \frac{3}{4}\left(\rho_{\Upsilon} + p_{\Upsilon}\right) \geq 0,
\end{align}
\begin{align}\label{bbb2}
  &\m{\Tcal^{(2)}} = 3p_{\Upsilon} - \rho_{\Upsilon},
\end{align}
showing explicitly that only the massive fields contribute to the
trace part of matter vacuum energy density. Within our level of
approximation, trace anomalies due to either matter interactions or
metric fluctuations are not taken into account.

{Finally, the averaged equation holding at the classical scale $\L$
and fixing the classical second order metric perturbation
$g^{(2)\cc}_{ab}$ is
\begin{equation}\label{2ndB}
  G^{(1)}_{ab}[g^{(2)\cc}]
  = 8\pi\left\langle\Tcf{}^{(2)}[\a] + \Tc^{(2)}[\Psi,\Upsilon]\right\rangle,
\end{equation}
where $\langle\Tcf{}^{(2)}[\a]\rangle$ is given in
\eqref{tst}-\eqref{ccc2} and
$\langle\Tc^{(2)}[\Psi,\Upsilon]\rangle$ in \eqref{tst2},
\eqref{bbb1} and \eqref{bbb2}.}

\subsection{Vacuum energy balance equation}

The classical, second order equation \eqref{2ndB} with no conformal
fluctuations ($\a = \text{const.} \Rightarrow \Sigma_{ab}\! =\! 0$)
would lead to $G^{(1)}_{ab}[g^{(2)\cc}] = 8\pi
\m{\Tc^{(2)}[\Psi,\Upsilon]}$. Given the structure of the vacuum,
averaged matter stress energy tensor in \eqref{tst2} we immediately
notice a problem connected to its traceless part. Indeed, even with
the adopted ultraviolet cutoff at the random scale $\l$, the energy
density $\rhom$ will attain a huge value and, \emph{more
interestingly}, will break the Lorentz invariance of vacuum at the
classical scale. Even in the idealized situation where the matter
fields were to be neglected, the traceless effective stress energy
tensor associated with graviton, included in $\Psi$, would still
cause the same problem.

The above conclusion can be avoided once the effects of the
nonlinear metric fluctuations are carefully taken into account. When
these are prescribed to include some amount of conformal
fluctuations, Lorentz invariance can be restored if the following
\emph{vacuum energy balance equation} holds
\begin{equation}\label{eq1}
  \rhocf + \rhom = 0.
\end{equation}
Then the only residual vacuum effect at the classical scale would
come in the form of an effective cosmological constant $\LM$ as
defined by
\begin{equation}\label{eq2}
\LM := -2\pi\frac{G}{c^4}\m{\Tcal^{(2)}[\Upsilon]} = \frac{8\pi
G}{c^4}\rho_M,
\end{equation}
where the matter related quantity $\rho_M$ has the dimensions of an
energy density and is defined as
\begin{equation}\label{rm}
    \rho_M := \frac{1}{4}(\rho_{\Upsilon}-3p_{\Upsilon}).
\end{equation}
For clarity we have explicitly displayed in \eqref{eq2} the factors
that convert a cosmological constant into the corresponding energy
density.

We will study the implications of equations \eqref{eq1} and
\eqref{eq2} in the following sections. They embody the choice of the
classical boundary conditions for the classical metric perturbation
$g^{(2)\cc}_{ab}$ in the sense that, once they can be shown to hold,
then the classical second order Einstein equation \eqref{2ndB} in
empty space reads
\begin{equation}\label{eecc}
      G^{(1)}_{ab}[g^{(2)\cc}] = -\LM\eta_{ab}.
\end{equation}
{This formally corresponds to the linear approximation of classical
Einstein equation with a cosmological constant $\Lambda$ and an
approximatively flat metric. Indeed, if $g_{ab} \approx \eta_{ab} +
h_{ab},$ with $\abs{h_{ab}} \ll 1$, then the linear terms of $G_{ab}
= - \Lambda g_{ab}$ would yield $G^{(1)}_{ab}[h] = -\Lambda
\eta_{ab}.$ This has precisely the same form as in \eqref{eecc},
provided that we identify $h_{ab} = g^{(2)\cc}_{ab}$ and $\Lambda =
\LM.$}

\subsection{Effective cosmological constant}\label{formal}

{The current estimated amount of observed vacuum energy in the
present epoch universe is} \cite{Carroll2001}
\begin{equation}\label{occ}
    \rho^{\text{\scriptsize{obs}}}_{\text{\scriptsize{vacuum}}}
    \approx 5\times 10^{-11}\text{ J m}^{-3} \approx 10^{-125}
    \rho_P,
\end{equation}
where $\rho_P \approx 10^{19}\text{ GeV}/L_P^3$ is the Planck energy
density. This value is deduced from the observed large scale
properties of the universe, including the approximatively flat
geometry, as suggested by CMB anisotropies \cite{boomerang}, and the
present, small, cosmic acceleration indicated by type Ia supernovae
observations \cite{Ia_supernovae}. Defining the Planck cosmological
constant as
\begin{equation}
\LP := \frac{8\pi G}{c^4}\rho_P,
\end{equation}
then \eqref{occ} implies the following observational constraint on
the present epoch cosmological constant
\begin{equation}\label{CCP}
0 \leq \Lobs \lesssim 10^{-125} \LP.
\end{equation}
The upper bound is extremely small, especially in comparison to the
Quantum Field Theory (QFT) theoretical expectations usually reported
in the literature. There it is commonly argued that, applying QFT
within the Standard Model and up to the Planck scale, the expected
cosmological constant would result to be of the order of $\LP$,
hence the famous discrepancy of about 120 orders of magnitude with
the observed value \cite{Carroll2001}.

Our random gravity framework also predicts that an effective
cosmological constant should indeed emerge due to vacuum energy
effects. However, at odds with the usual claims, only the massive
fields are expected to give a contribution. A necessary requirement
for the observational constraint not to be violated is that $\LM$ be
positive. This implies that massive matter fields in the vacuum
state should satisfy the following condition
\begin{equation}\label{rhom}
\rho_M = \frac{\LM c^4}{8\pi G} = \frac{1}{4}(\rho_{\Upsilon} -
3p_{\Upsilon}) \geq 0.
\end{equation}
In appendix \ref{A1} it is shown that, in a first approximation that
treats matter fields in their vacuum state as a collection of
non-interacting harmonic oscillators, $\rho_M$ is positive and
proportional to a squared mass parameter $M^2$, defined in equation
\eqref{ap18} as a weighted average of the squared masses of all
particles comprised in $\Upsilon$. Specifically, we obtain in \eqref{ap19}
and \eqref{ap20} that
\begin{equation}\label{rhoMval}
\rho_M =
\frac{N_{\Upsilon}\rho_P}{8\lambda^2}\left(\frac{M}{M_P}\right)^2
\end{equation}
\begin{equation}\label{LMval}
\LM =
\frac{N_{\Upsilon}\LP}{8\lambda^2}\left(\frac{M}{M_P}\right)^2.
\end{equation}
The above clearly shows that $ \rho_M \propto
\frac{\rho_P}{\lambda^2} \left(\frac{M}{M_P}\right)^2, $ where $M_P
\approx 10^{19}$ GeV is the Planck mass. This value can still be
large and formally diverges as $\lambda$ decreases. If a Planck
cutoff scale $\lambda \sim 1$ were to be applied applied as per the
``usual practice'', we would end up with an effective cosmological
constant $\LM$ much smaller than the Planck value $\LP$, roughly by
a factor of $({M}/{M_P})^2$. Estimating this using the maximum mass
scale of the Standard Model, namely $M = M_H \approx 100$ GeV, then
the reduction factor is $({M_H}/{M_P})^2 \sim 10^{-34}$ and the
resulting actual value for $\LM$ still would exceed the
observational value $\Lobs$ by about 90 orders of magnitude. However
our present result, indicating that the cosmological constant seems
to be related to the matter fields masses, opens up new a prospect
for considerations that will be discussed in the final section of
this paper.

\section{Analysis of the balance equation}

\subsection{Lorentz invariance of vacuum and conformal fluctuations}

Turning now to the issue of Lorentz invariance, we can use
\eqref{ccc1} and the balance equation \eqref{eq1} to get
\begin{equation}\label{eq1B}
    3\m{\a_{,0}^{\,2}} + \m{\abs{\nabla\a}^2} =
    8\pi\rhom.
\end{equation}
In principle, this equation can always be satisfied and comes in the
form of a statistical condition for the conformal fluctuations $\a$.
It requires that their ``intensity'' be precisely adjusted so to
balance the traceless part of the vacuum energy contribution due to
all the other fluctuating fields. At the present stage this
requirement is imposed ``from the outside'', on the basis that
Lorentz invariance of vacuum should hold at the classical scale.
However we would expect that a high energy, underlying quantum
theory of gravity that incorporates conformal fluctuations as an
essential ingredient, could lead to a more detailed and fundamental
``internal'' explanation for the balancing mechanism to hold. Some
preliminary but promising indications along these lines are given in
\cite{Wang2005,Wang2005b}, where the canonical formulation of
gravity is enriched to include conformal symmetry in both the
geometrodynamics and connection variables approaches.

\subsection{Characterizing $\a$ using the wave equation}

The conformal metric fluctuations provide a mechanism to restore
Lorentz invariance of vacuum at the classical scale provided they
have suitable statistical properties as expressed by the averages
$\m{\a_{,0}^{\,2}}$ and $\m{\abs{\nabla\a}^2}$. We recall at this
point that since the conformal field $\a$ appears in our present
formalism as an \emph{external field} that we can prescribe at will,
we are free to assign it in such a way that it satisfies an extra
dynamical constraint without affecting the Einstein equation. To be
guided in our choice we notice that the second order effective
stress energy tensor in \eqref{tr} is traceless and depends solely
on the first derivatives of $\a$. If it had to have a trace term,
the simplest and most natural combination containing first
derivatives of $\a$ that would serve the purpose is
\begin{equation}
\a^{,c} \a_{,c} \equiv -\a_{,0}^{\,2} + \abs{\nabla \a}^2.
\end{equation}
Therefore a simple and reasonable statistical condition to impose on
the conformal fluctuations appears quite naturally to be
\begin{equation}\label{stat3}
\m{\a^{,c} \a_{,c}} \equiv -\m{\a_{,0}^{\,2}} + \m{\abs{\nabla
\a}^2}=0.
\end{equation}

Then a natural choice for the extra dynamical constraint is
suggested by the fact that (i) $\a$ is a scalar field and (ii) it is
expected to satisfy the statistical relation \eqref{stat3}. It is
\begin{equation}\label{KG}
    \pd^c\pd_c \a = 0,
\end{equation}
i.e. the simple wave equation on Minkowski spacetime for a massless
scalar field. Indeed it is readily verified that, provided the
statistical properties of $\a$ are stationary, equation \eqref{KG}
guarantees the condition $\m{\a_{,0}^{\,2}} - \m{\abs{\nabla \a}^2}
= 0$ to hold. Our choice is further motivated by the fact that the
usual stress-energy tensor for a massless scalar field $\phi$ on a
flat spacetime has the form
\begin{equation}\label{KGT}
    \TKG := \left(\phi_a\phi_b -
\frac{1}{4}\eta_{ab}\phi^c\phi_c\right) +
\frac{1}{4}\eta_{ab}\left(-\phi^c\phi_c\right).
\end{equation}
Provided $\m{\phi^c\phi_c} = 0$ and modulo the irrelevant rescaling
factor of $1/2\pi$, its average matches precisely the expression in
\eqref{tr}, but with the negative sign.

Once the wave equation for $\a$ is introduced, it can be solved with
Boyer's type (random) boundary condition and the overall (classical)
statistical amplitudes set in such a way that the balance equation
is satisfied. It is precisely in this sense that the conformal
random fluctuations are independent of graviton's and those of other
matter fields. However, their overall amount is fixed at the
\emph{classical} level by the requirement of Lorentz invariance
alone. Using \eqref{stat3} the balance equation takes the very
simple form
\begin{equation}\label{eq1C}
    \m{\abs{\nabla\a}^2} =
    2\pi\rhom.
\end{equation}

An important physical justification of the wave equation \eqref{KG}
for $alpha$ is that its solution admits a Lorentz invariant spectral
density suitable for describing vacuum fluctuations. Further, the
form of this spectral density allows the effective stress-energy
tensor of the conformal field to regularize the vacuum energy of
matter as detailed below.

The formal analogy between the stress-energy tensors \eqref{KGT} and
\eqref{tr} and equation \eqref{eq1C} above suggest that we introduce
a rescaled conformal field as
\begin{equation}\label{rcf}
    \aa := \frac{\a}{\sqrt{2\pi}}.
\end{equation}
With this notation the balance equation reads
\begin{equation}\label{eq1D}
    \m{\abs{\nabla\aa}^2} = \rhom.
\end{equation}

\subsection{Isotropic power spectrum of the conformal fluctuations}

The wave equation implies the usual dispersion relation $\omega^2 =
k^2$ relating frequency $\omega$ and wave number magnitude
$k:=\abs{\kv}$. The conformal field can accordingly be expressed in
Fourier components whose corresponding \emph{power spectral density}
$S_{\aa}(\kv)$ can be defined in a statistical sense
\cite{Bonifacio2008}. Since vacuum is isotropic at the classical
scale we have accordingly $S_{\aa} = S_{\aa}(k)$. It is then
straightforward to show that
\begin{equation}\label{Ag2}
   \m{\aa^2} = \frac{1}{(2\pi)^3}\int d^3 k\, S_{\aa}(k),
\end{equation}
\begin{equation}\label{Ag}
   \m{\abs{\nabla \aa}^2} = \frac{1}{(2\pi)^3}\int d^3 k\, k^2 S_{\aa}(k),
\end{equation}
\begin{equation}\label{Ag1}
   \m{\aa_{,0}^2} = \frac{1}{(2\pi)^3}\int d^3 k\, \omega^2(k) S_{\aa}(k),
\end{equation}
which, given the dispersion relation, are of course compatible with
\eqref{stat3}.

Since $\int d^3 k \equiv 4\pi\int dk k^2$ these imply an integration
over the wavelengths of the conformal fluctuations. In the present
context we shall adopt a UV, i.e. high frequency, cutoff. We take
the corresponding minimum wavelength to be equal to the random scale
$\l = \lambda L_P$. This is because our random formalism based on
the concept of classical fluctuating fields breaks down for scales
shorter that $\l$. We thus have the wave number cutoff
\begin{equation}\label{cut}
    \kcut:=\frac{2\pi}{\lambda L_P}.
\end{equation}
From the physical view point the dimensionless cutoff parameter
$\lambda$ should be in the range of $\lambda \gtrsim 10^0$, below
which the quantum spacetime effects are expected to be important.
However, as far as Lorentz invariance of vacuum is concerned, it can
also be thought as a regularization parameter whose precise value is
not critical. This is so because the same UV cutoff is to be applied
to the calculation of the matter related energy density and
pressure. As a result, although the resulting expressions for
$\m{\abs{\nabla\aa}^2}$ and $\rhom$ formally diverge as $\lambda
\rightarrow 0$, their ratio and the related balancing mechanism turn
out to be cutoff independent.

The statistical quantities defined above must be invariant under
Lorentz transformations. To this end we take
\begin{equation}\label{ps}
    S_{\aa}(k):= \frac{S_0 \hbar G}{c^2}\frac{1}{\omega(k)},
\end{equation}
where, being $\aa$ dimensionless, the combination of $\hbar$, $G$
and $c$ gives the right dimensions for a power spectrum (i.e.
$L^3$), and $S_0$ is a dimensionless constant that controls the
overall `normalization'. Our choice is motivated by the fact that
this yields the Lorentz invariant measure $d^3 k/\omega(k)$
\cite{ryder} and it implies an {energy spectrum} $\rho(\omega)
\propto \omega^3,$ which is shown by Boyer (see ref. below) to be
the only Lorentz invariant vacuum energy spectrum for a massless
field. Of course Lorentz invariance is preserved provided the cutoff
$\kcut$ is given by the \emph{same number} for all inertial
observers, as discussed in details by Boyer \cite{Boyer1969}.

\subsection{Explicit solution of the balance equation}

Thanks to equations \eqref{Ag2}-\eqref{Ag1} all the averages
involving the conformal fluctuations simply depend upon the spectral
parameter $S_0$ and, of course, the cutoff parameter $\lambda$.
Although it functions here as a regularization parameter, we recall
that a specific value for $\lambda$ may mark an effective transition
scale from the random to the purely quantum domain. In principle,
this effective cutoff value could be estimated, for example, through
high precision measurement of the decoherence suffered by massive
quantum particles propagating in vacuum \cite{PowerPercival2000,
Wang2006, Gerlich2007, gauge2008}. This possibility will be further investigated in the
follow up paper \cite{Bonifacio2008}.

Restoring full physical units by inserting back the constants $c,
\hbar$ and $G$, equation \eqref{eq1D} is
\begin{equation}\label{x1}
\rho_P L_P^2 \m{\abs{\nabla\aa}^2} = \rhom
\end{equation}
where $L_P$ is the Planck length. The dispersion relation with
restored physical constants is $ \omega(k) = ck $ and the integrals
\eqref{Ag2}-\eqref{Ag} yield
\begin{equation}\label{i1}
\m{\aa^2} = \frac{S_0}{\lambda^2},
\end{equation}
\begin{equation}\label{i2}
\m{\abs{\nabla\aa}^2} = \frac{2\pi^2}{L_P^2}\frac{S_0}{\lambda^4}.
\end{equation}
Substituting \eqref{i2} into \eqref{x1} we have immediately
\begin{equation}\label{sol5}
    S_0 = \frac{\lambda^4}{2\pi^2}\frac{\rhom}{\rho_P}.
\end{equation}
As discussed above, the cutoff parameter $\lambda$ dependence is
only apparent. The matter fields ``traceless'' energy density
$\rhom$ is calculated in appendix \ref{A1} as
\begin{equation}
\rhom = \frac{\pi^2 N \rho_P}{\lambda^4},
\end{equation}
where $N$ is the total number of independent matter fields modes.
Substituting into \eqref{sol5}, this leads to the cutoff independent
result
\begin{equation}\label{sol6}
    S_0 = \frac{N}{2},
\end{equation}
representing precisely the power spectrum normalization that is
needed to balance matter fields, traceless vacuum energy density as
described by $\rhom$ and restore Lorentz invariance.

Plugging this value back into equation \eqref{i1}, the average
conformal fluctuations squared amplitude follows as
\begin{equation}
\m{\a^2} = \frac{N}{2\lambda^2}.
\end{equation}

\section{Interpretation and discussion}

If vacuum energy has any physical reality then it should act as a
source of gravity according to the Einstein equation. Quantum field
theory predicts a large amount of vacuum energy that may lead to a
Planck size cosmological constant. However current observations of a
small cosmic acceleration suggest that the overall amount of vacuum
energy in the universe must be many order of magnitudes smaller than
predicted by the standard approach.

We have provided a general random gravity framework in which
spacetime vacuum fluctuations are described by means of a stochastic
metric satisfying the ordinary Einstein field equation. As it was
shown by Boyer for the electromagnetic field, this is equivalent to
specifying appropriate fluctuating boundary conditions. The
resulting fluctuating classical fields mimic quantum fluctuations,
in the sense that a variety of related phenomena can be accounted
for. In this work we assume this to be valid in the case of
spacetime fluctuations, up to some energy scale which is set by the
random scale close to the Planck scale. These are described by
fluctuating metric perturbations defined on a flat background
metric. The resulting random gravity theory is built so as to
provide a description of the apparent empty spacetime from the
random scale up to some large scale $\up$, where the deviations of
the background from a flat Minkowski start to be significant. To
this end, a spacetime averaging procedure is employed to describe
the passage from the fluctuating metric to a classical metric that
describes spacetime beyond some classical scale $\L \gg \l$.

It turns out that, in order for the resulting theory to be
compatible with the Lorentz invariance of vacuum at the classical
scale, the fluctuating metric must execute conformal fluctuations
spontaneously. This yields a metric of the form $\gamma_{ab} =
\exp(2\a)\g$. This equation must properly be interpreted as a simple
statement about the two metric tensors $\gc$ and $\g$ being
\emph{conformally related}. The issue of which of them actually
``contains'' the conformal fluctuations is an interesting one. In
the present framework this questions is partially answered by
prescribing the fluctuations in $\a$ to be independent from the
fluctuations in the first order metric perturbation
$\beta_{ab}=\gamma_{ab}^{(1)}$, supposed to mimic graviton vacuum
fluctuations. Once this is done, $\a$ appears to be related to the
trace of $q_{ab}=g^{(1)}_{ab}$, i.e. the first order perturbation of
the effective metric $\g$. In principle we can cast $\beta_{ab}$
into a {\small{\texttt{TT}}} gauge, so that $\beta = 0$ and $\a
\equiv -q/8$, where $\beta$ and $q$ denote the traces. In this sense
we can claim that conformal fluctuations are actually ``contained'',
\emph{to first order}, within the effective metric $\g$ but
\emph{not} within the metric $\gamma_{ab}$. However, conformal
fluctuations propagate through nonlinearly coupling in such a way
that $\a$ should affect the higher order metric perturbations of
\emph{both} metric tensors $\g$ and $\gc$.

In the first order approximation $\a$ does not affect graviton
fluctuations and it just acts as a linear source of random noise for
$q_{ab}$. This happens in such a way that, at the classical scale,
the classical tensors $\m{\beta_{ab}}$ and $\m{q_{ab}}$ coincide and
satisfy the usual linearized equation for GWs with no extra sources.
In the second order the conformal fluctuations described by $\a$
induce a traceless, negative definite, effective stress energy
tensor which acts as a source for the effective metric $\g$. Under
some circumstances, this can counterbalance the predominant part of
vacuum energy due to matter fields and coming from the traceless
part of their associated vacuum stress energy tensor. Without such a
counterbalancing mechanism, the associated stress-energy tensor
attains a large, close to Planckian magnitude, thereby breaking
Lorentz invariance, even with an UV cutoff set at the random scale
$\l = \lambda L_P$.

We showed that Lorentz invariance can be restored by the conformal
fluctuations by means of a simple balancing mechanism, in such a way
that the large, traceless, matter fields contribution is canceled
out by the negative, conformal contribution. This regularization
process fixes the averaged conformal fluctuation squared amplitude
unambiguously once these are chosen to have a Lorentz invariant
power spectrum and to satisfy certain specific statistical
properties. These are shown to be compatible with a simple massless
wave equation ruling the conformal fluctuations dynamics. This extra
dynamical constraint is consistent with the Einstein equation since
the conformal field is originally prescribed as an external field
and, in principle, can be assigned arbitrarily. The main choice
related to its statistical independence with graviton fluctuations
at the random level, together with the Lorentz invariance
requirement at the classical level, effectively attach physical
meaning to the conformal field fluctuations and fix their
statistical properties.

The dimensionless cutoff parameter $\lambda$ introduced in the
process is likely to mark the transition from the random domain to
the purely quantum domain at the Planck scale. In this sense it sets
the small scale breakdown limit to our current theoretical approach.
However it is very interesting to note that, in a similar way as to
what happens with the Casimir effect, the spectral density of the
conformal fluctuations that results from the balance mechanism is
cutoff independent. This is a strong result, ``almost'' entirely
based upon general relativity alone. The ``almost'' simply referring
to the fact that, to implement the regularization process, it has
been necessary to modify the classical boundary conditions and enter
into the random fields domain.

The final outcome of the regularization process is (i) a Lorentz
invariant ``empty'' spacetime at all classical scales and (ii) an
effective cosmological constant $\LM$, emerging as a second order
nonlinear effect due to the trace part of the matter fields vacuum
stress-energy tensor. The effective cosmological constant thus
appears as the only remnant effect, built up from the quantum
vacuum, which could in principle have an effect at a large scale. At
odds with what commonly stated, the massless fields do not seem to
play any role in contributing to the effective cosmological
constant, at least within the present second order nonlinear
approach. Our final equation for $\LM$ shows that it is proportional
to a squared mass parameter $M^2$, obtained as a weighted average
over all massive fields.

Even when a usual cutoff at the Planck scale (i.e. $\lambda \sim 1$)
is applied, our formula yields a leftover, effective cosmological
constant which is roughly 30 orders of magnitude lower than the
usual Planckian value, though still much larger than the almost
vanishing value value $\Lobs \approx (8\pi G/c^4)10^{-125}\rho_P$
suggested by cosmological observations. Our current result could
however be reinterpreted under a slightly different and interesting
perspective: by identifying $\LM = \Lobs$, we get an equation for
the matter field averaged effective mass
$M_{\lambda}=M/(M_P/\lambda)$, namely
\begin{equation}
    M_{\lambda} \approx \sqrt{\frac{\Lobs}{(\LP/\lambda^4)}} \equiv
    \sqrt{\frac{\rho^{\text{\scriptsize{obs}}}_{\text{\scriptsize{vacuum}}}}{(\rho_P/\lambda^4)}}.
\end{equation}
Once the numbers are plugged in this yields the result $M \approx
\lambda M_P \sqrt{10^{-125}} \approx \lambda \times 10^{-44}$ GeV.
This result shows that, for any realistic value of the cutoff
parameter, \emph{including} the Planckian value $\lambda\approx 1$,
the average matter fields effective mass matching the currently
observed cosmological constant value is basically vanishing. This
result is interesting since it links to some scenarios discussed in
the literature and suggesting that, at the most fundamental level,
matter fields \emph{bare masses} may in fact be zero
\cite{Wilczek1999, Meissner2007}. The observed particle masses would
appear as a result of interactions. One example would be a Higgs,
complex scalar field based interaction \cite{Higgs1964, Quigg2007,
Okada2007}, but other possibilities are being considered
\cite{Meissner2007, Wilczek2005}.

The particular approach used in this paper, where matter
interactions are not included, could be relevant in this perspective
if we interpret the effective mass $M$ as the average bare mass of
the matter fields in their vacuum state. If this is zero as
suggested by some authors and, we add, if the actual mass generating
mechanisms are not effective in the vacuum state, then a possible
solution to the riddle of the cosmological constant problem could be
``almost'' at hand. We say ``almost'' since any sound statement
would first require a solid and deep understanding of the nature of
vacuum and, especially, of the meaning of mass and how it comes to
be. At this stage we point out however that, with the conformal
fluctuations as a possible mechanism to yield Lorentz invariance and
neutralize most part of the matter fields vacuum energy density, the
possible connection between cosmological constant and effective
(bare) matter fields mass is very interesting.

A widely accepted present view is this: While physics may well
restore conformal invariance at a unified high energy scale where
all fundamental particles become massless, these particles
nonetheless gain an effective mass by settling into a local minimum
of the Higgs potential (or any other mass generating mechanism) at
low energies. This implies that the effective cosmological constant
$\LM$ is built from nonzero mass up to the Higgs (or equivalent)
energy scale, say $M_H \sim 100$GeV, corresponding to an effective
cutoff at $\lambda\sim M_P/M_H$ in \eqref{LMval}. This equation,
together with $M\sim M_H$ then yields the estimated  effective
cosmological constant to be
\begin{equation}\label{LH}
\LM \sim \LP\left(\frac{M_H}{M_P}\right)^4 \sim 10^{-68}\LP\sim
10^{57}\Lobs
\end{equation}
which still exceeds the observational value by about 57 orders of
magnitude. However, this estimate assumes a simplistic zero value of
the local minimum of the Higgs potential. It has been suggested
that, it is the local maximum, where the Higgs field vanishes, that
should be zero \cite{Weinberg1989}. While other ``calibrations'' may
exist, it is interesting to note that if this is true, then the
equilibrium potential energy of the Higgs field would amount to an
additional effective cosmological constant, that is (roughly within
the same order of magnitude) equal and opposite to $\LM$ given in
\eqref{LH}. Whether or not this possible cancelation would result in
a net effective cosmological constant even closer to the
observational value is an intriguing issue for future investigation,
where further interaction effects ought to be considered as well.

We conclude with the following comment. We have shown that general
relativity permits spontaneous fluctuations of metric, and the
conformal fluctuations in particular. These bear dynamical
consequences on the evolution of the averaged metric. Once the
spontaneously  fluctuating conformal field is set to regulate the
net amount of vacuum energy, it appears to have somehow a life on
its own. In this sense the conformal field is as real as the vacuum
energy. One then wonders why we never have  detected it. In this
respect we notice that even standard gravitational waves are so weak
that, to date, still fail to be detected. Hopefully projects such as
LIGO and LISA will allow a direct unambiguous detection of GWs. In
the case of vacuum conformal fluctuations, we are talking about tiny
modulations of space time that happen at the random scale, i.e. $
\gtrsim 10^{-35}$ m. These modulations average out at the classical
scale. This makes them ideally very difficult to detect directly,
even though there is a possibility that matter wave interferometry
could provide experimental evidence in measuring the related
decoherence on massive, non-relativistic quantum particles
\cite{Gerlich2007, gauge2008}. At this stage, it is clear however
that, once the role of the conformal fluctuations in regularizing
the vacuum energy of the universe has been clarified, an indirect
evidence of their reality could be, after all, the fact that
observations seem to hint at a vacuum which is compatible with
Lorentz invariance at the classical scale and a positive, small,
cosmological constant.

The inclusion of spontaneous conformal spacetime fluctuations as a
fundamental ingredient of the quantum vacuum seems to yield
promising results. More efforts will be needed to clarify their
deeper, quantum, meaning and cast further light upon the
significance and the mutual relation between, quantum vacuum,
cosmological constant, as well as the generation mechanism and the
ultimate nature of mass.

\newpage

\begin{acknowledgments}

The authors are most grateful to G. Amelino-Camelia (Rome), M. Arndt
(Vienna), J. Brown (Glasgow), K. Hornberger (Munich), C.
L\"ammerzahl (ZARM Bremen), J. Reid (Aberdeen) and T. Sumner (IC
London) for stimulating discussions. PB thanks the University of
Aberdeen for a Sixth Century Ph.D. Studentship Award. The research
is supported by the STFC Centre for Fundamental Physics.

\end{acknowledgments}

\appendix

\section{Energy density and pressure of matter fields using the free field approximation}\label{A1}

Here we estimate the matter fields related quantities $\rhom$ and
$\rho_M$ in full physical units. From their definitions in
\eqref{bbb1} and \eqref{rhom}, we need evaluate the three quantities
$\rho_{\Psi},$ $\rho_{\Upsilon}$ and $p_{\Upsilon}.$ These can in
principle be obtained with quantum field theory. However this is not
a trivial task. Within the Standard Model, vacuum energy density is
estimated to be given by, at least, three main contributions
\cite{Rugh02}: (i) vacuum zero-point energy plus virtual particles
fluctuations, (ii) QCD gluon and quark condensates (iii) Higgs
field.

Through the following calculation we want to get a first
approximation estimate, by neglecting fields interactions and by
describing the free-field configuration as a collection of decoupled
harmonic oscillators of frequency
\begin{equation}\label{wkm}
\omega_{k}=\sqrt{c^2k^2 +m^2c^4},
\end{equation}
where $k$ is the norm of the spatial wave vector $\kv$ and $m$ is
some \emph{effective mass} of the field quanta under examination.
This is quite accurate for the EM field but it is likely to be quite
a crude approximation for e.g. the QCD sector. By doing so we are
neglecting higher order contributions to the vacuum energy density
as well as the nonlinear and strong coupling effects of QCD and a
detailed treatment of the Higgs fields. Nonetheless we expect to
obtain a meaningful lower bound estimate to the vacuum energy.

We consider an arbitrary matter field of effective mass $m$ within
the Standard Model. The field is comprised in $\Psi$ or $\Upsilon$
depending on whether it is $m=0$ or $m\neq 0$. We calculate a lower
bound to the associated vacuum energy density by estimating the
overall energy density resulting from the summation of the zero
point energies of all frequency modes, up to the random scale cutoff
set by $\l = \lambda L_P.$

We model each independent field component as a scalar field $\phi$
with mass $m$ and associated Klein-Gordon stress-energy tensor
\begin{equation}\label{strs}
\TKG[\phi]=\phi_a\phi_b-\frac12\eta_{ab}
\left(\phi^c\phi_c+\frac{m^2c^2}{\hbar^2}\phi^2\right).
\end{equation}
From this we see that $\phi^2$ has the dimension of force. Regarding
$\phi$ as in its zero-point fluctuating state, we shall assume it to
be stationary over spacetime and having an isotropic spectral
density $\Sphi(k)$, so that the mean squared field is given by
\begin{equation}\label{mphi}
\m{\phi^2}= \frac{1}{(2\pi)^3}\int d^3 k\,\Sphi(k) .
\end{equation}
Furthermore, the mean squared time derivative
$\phi_{,t}=c\phi_{,0}$ satisfies
\begin{equation}\label{mphid}
\mm{\phi_{,t}^{\;2}}= \frac{1}{(2\pi)^3}\int d^3 k\,\omega_k^2
\Sphi(k)
\end{equation}
and the mean squared gradient $(\nabla\phi)_i:=\phi_{,i}$
for $i=1,2,3$ satisfies
\begin{equation}\label{mphip}
\m{\abs{\nabla\phi}^2}= \frac{1}{(2\pi)^3}\int d^3 k\,k^2 \Sphi(k).
\end{equation}

It follows from \eqref{wkm}, \eqref{mphi}, \eqref{mphid},
\eqref{mphip} that
\begin{equation}\label{rel1}
\m{\phi^a\phi_a}=-\frac1{c^2}\mm{\phi_{,t}^{\;2}}+\m{\abs{\nabla\phi}^2}
=-\frac{m^2c^2}{\hbar^2}\m{\abs{\phi}^2}.
\end{equation}
Using \eqref{mphid}, \eqref{mphip} and \eqref{rel1} we see that the
mean stress-energy tensor given by \eqref{strs} becomes
\begin{equation}\label{strsB}
\m{\TKG}=\m{\phi_a\phi_b},
\end{equation}
which yields the effective energy density
\begin{equation}\label{phirho1}
\rho=\m{T_{00}}=\frac1{c^2}\mm{\phi_{,t}^{\;2}}
=\frac{1}{(2\pi)^3}\int d^3 k\,\frac{\omega_k^2}{c^2} \Sphi(k)
\end{equation}
and the effective pressure
\begin{equation}\label{phiP1}
p= \frac13\mm{T^i{}_i} =\frac13\m{\abs{\nabla\phi}^2}
=\frac{1}{3(2\pi)^3}\int d^3 k\,k^2 \Sphi(k).
\end{equation}
A useful combination follows from \eqref{wkm}, \eqref{phirho1} and
\eqref{phiP1} as
\begin{equation}\label{phirhoP2}
\rho-3p=\frac{1}{(2\pi)^3}\frac{m^2c^2}{\hbar^2}\int d^3 k\,
\Sphi(k).
\end{equation}

The spectral density $\Sphi(k)$ itself can be determined through the
well-known zero point energy density expression
\begin{equation}\label{phirho2}
\rho=\frac{1}{(2\pi)^3}\int d^3 k\,\frac{\hbar\omega_k}{2},
\end{equation}
that adds up contributions from all wave modes of $\phi$. Comparing
\eqref{phirho1} and \eqref{phirho2} we see that the spectral density
$\Sphi(k)$ takes the Lorentz invariant form
\begin{equation}\label{phiSDk}
\Sphi(k)=\frac{\hbar c^2}{2\omega_k}.
\end{equation}

Substituting \eqref{wkm} and \eqref{phiSDk} into \eqref{phiP1} and
\eqref{phirhoP2} and integrating using $\int d^3 k = 4\pi\int dk
k^2$ up to the cutoff value $k_\lambda=\frac{2\pi}{\lambda L_P}$, we
obtain
\begin{equation}\label{phirho4}
\rho =\frac{\hbar}{4\pi^2}\int_0^{k_\lambda}d k\,k^2
\sqrt{c^2k^2+\frac{m^2c^4}{\hbar^2}},
\end{equation}
and
\begin{equation}\label{phiP4}
p =\frac{\hbar c^2}{12\pi^2}\int_0^{k_\lambda}d k\,
{k^4}\Big/{\sqrt{c^2k^2+\frac{m^2c^4}{\hbar^2}}}.
\end{equation}
Introducing the dimensionless variable $y:= \frac{k\hbar}{mc}$, we
can rewrite \eqref{phirho4} and \eqref{phiP4} as
\begin{equation}\label{phirho5}
\rho = \frac{\rho_P}{4\pi^2}\left(\frac{m_\lambda}{\lambda}\right)^4
 \int_0^{\frac{2\pi}{m_\lambda}}\!\!\!dy\,y^2\sqrt{1+y^2},
\end{equation}
and
\begin{equation}\label{phiP5}
p =\frac{\rho_P}{12\pi^2}\left(\frac{m_\lambda}{\lambda}\right)^4
\int_0^{\frac{2\pi}{m_\lambda}}d y\, \frac{y^4}{\sqrt{1+y^2}},
\end{equation}
in terms of the Planck energy density $\rho_P$ and the effective mass of
the field in units of $M_P / \lambda$, i.e. $ m_\lambda := \frac{m}{(M_P/\lambda)}$.

For $m_\lambda \ll 1$ we can approximate \eqref{phirho5},
\eqref{phiP5} by
\begin{equation}\label{phirho6}
\rho = \frac{\pi^2 \rho_P}{\lambda^4} +
    \frac{\rho_P}{4\lambda^2}\left(\frac{m}{M_P}\right)^2
\end{equation}
and
\begin{equation}\label{phiP6}
p = \frac{\pi^2 \rho_P}{3\lambda^4} -
    \frac{\rho_P}{12\lambda^2}\left(\frac{m}{M_P}\right)^2,
\end{equation}
up to adding $\mathcal{O}(m_\lambda^4)$ terms. This approximation is
physically well justified. The heaviest particles in the Standard
Model are the quark top, with $m_t \approx 173$ GeV, the weak bosons
$W^{\pm}$, with $m_{W^{\pm}} \approx 80$ GeV and the $Z$ boson with
$m_Z \approx 91$ GeV. All the other particles have $m \lesssim 1$
GeV. Being the Planck mass of the order of $10^{19}$ GeV, it will be
$m_\lambda \ll 1$ as long as the cutoff parameter satisfies $\lambda
\lesssim 10^{15}$. This is a safe upper bound, as the stochastic
classical conformal fluctuations are expected to have a cutoff which
should not exceed $\lambda \approx 10^2 - 10^5$ \cite{Wang2006,
Bonifacio2008}.

Within this approximation, by subtracting \eqref{phirho6} and
\eqref{phiP6} we get
\begin{equation}\label{phirhosubP}
\rho - 3p = \frac{\rho_P}{2\lambda^2}\left(\frac{m}{M_P}\right)^2.
\end{equation}
This relation can also be obtained directly from the right hand side
of \eqref{phirhoP2} by following through the steps leading from
\eqref{phirho4} to \eqref{phiP6}.

The expressions \eqref{phirho6} and \eqref{phiP6} are the sums of
two terms, one of which depends upon the effective mass of the
field. Collecting the contribution from all massless fields we have
the corresponding energy density $c^2\rho_{\Psi}$ and pressure
$p_{\Psi}$ given by
\begin{equation}\label{ap14}
\rho_{\Psi} = 3p_{\Psi} = \frac{N_{\Psi} \pi^2 \rho_P}{\lambda^4},
\end{equation}
where $N_{\Psi}$ is the total number of independent components for
{all} massless fields.

In the massive case, the total energy density $\rho_{\Upsilon}$
follows from all contributing fields with individual effective mass
$m_j$ and number of independent components $N_{\Upsilon_j}$. Using
\eqref{phirho6} we have
\begin{equation}\label{ap14b}
\rho_{\Upsilon} = \frac{N_\Upsilon \pi^2\rho_P}{\lambda^4} +
    \frac{N_\Upsilon^2
    \rho_P}{4\lambda^2}\left(\frac{M}{M_P}\right)^2,
\end{equation}
where $N_{\Upsilon} := \sum_{j}N_{\Upsilon_j}$ is the total
number of independent components of the massive fields and
\begin{equation}\label{ap18}
M^2:=\frac{\sum_{j} N_{\Upsilon_j} m_j^2}{N_{\Upsilon}}
\end{equation}
is the weighted average squared effective mass. The total pressure
of the massive matter fields then follows from
\eqref{phiP6} to be
\begin{equation}\label{ap16}
p_{\Upsilon} = \frac{N_\Upsilon \pi^2\rho_P}{3\lambda^4} -
    \frac{N_\Upsilon
    \rho_P}{12\lambda^2}\left(\frac{M}{M_P}\right)^2.
\end{equation}
We therefore have
\begin{equation}\label{ap19}
\rho_M := \frac{1}{4}(\rho_{\Upsilon}-3p_{\Upsilon}) =
\frac{N_{\Upsilon}\rho_P}{8\lambda^2}\left(\frac{M}{M_P}\right)^2,
\end{equation}
corresponding to the effective cosmological constant
\begin{equation}\label{ap20}
\LM =
\frac{N_{\Upsilon}\LP}{8\lambda^2}\left(\frac{M}{M_P}\right)^2.
\end{equation}

Finally the leading contribution of $\rhom$ can be obtained by
substituting \eqref{ap14}, \eqref{ap14b} and \eqref{ap16} into
\eqref{bbb1} to be
\begin{equation}\label{ap21}
\rhom = \frac{\pi^2 N \rho_P}{\lambda^4},
\end{equation}
up to adding $\mathcal{O}(M_\lambda^2)$ terms, where
\begin{equation}
    M_\lambda:= \frac{M}{(M_P/\lambda)},
\end{equation}
and
\begin{equation}\label{apN}
N := N_{\Psi} + N_{\Upsilon}
\end{equation}
is the total number of independent components for massless and
massive fields. The ratio between the traceless part and effective
cosmological constant energy densities for the matter fields scales
as
\begin{equation}
    \frac{\rho_M}{\rhom} \propto M_{\lambda}^2 \ll 1,
\end{equation}
where the inequality holds within the Standard Model, with $M
\approx 10^2$ GeV, and for any realistic value of the cutoff
parameter.

\section{Some technical derivations}\label{A2}

The linearized Einstein tensor has the explicit form
\cite{Flanagan2005}
\begin{align}\label{l1}
    G^{(1)}_{ab}[h]:=&\frac{1}{2}\pd^c\pd_b h_{ac} +
    \frac{1}{2}\pd^c\pd_a h_{bc} - \frac{1}{2}\pd^c\pd_c h_{ab}\nonumber\\
    &-\frac{1}{2}\pd_a\pd_b h - \frac{1}{2}\eta_{ab}
    \left( \pd^c\pd^d h_{cd} - \pd^c\pd_c h \right),
\end{align}
where $h:= \eta^{ab}h_{ab}$ denotes the trace. It is straightforward
to verify that, for $h_{ab} \equiv 2\a\eta_{ab}$ this yields
\begin{align}\label{l1B}
    G^{(1)}_{ab}[2\a\eta_{ab}]=2\eta_{ab}\pd^c\pd_c\a -
    2\pd_a\pd_b\a,
\end{align}
as used in \eqref{yyy}.

Turning now to the calculation of $\m{\Tcf{}^{(2)}}$ in
\eqref{mtcf}, the full explicit expression of
$-\m{G^{(2)}_{ab}}/8\pi$ can be simplified to yield the following
nonlinear differential operator \cite{Flanagan2005}
\begin{align}\label{zzz}
    &-\frac{1}{8\pi}\m{G^{(2)}_{ab}[h]} = \frac{1}{32\pi}\times\nonumber\\
    &\m{\pd_a \bar{h}_{cd} \pd_b \bar{h}^{cd} -
    \frac{1}{2}\pd_a \bar{h} \pd_b \bar{h}
    -\pd_a \bar{h}_{bc} \pd_d \bar{h}^{cd} -
    \pd_b \bar{h}_{ac} \pd_d \bar{h}^{cd}},
\end{align}
where
\begin{equation}
    \bar{h}_{ab}:= h_{ab} - \frac{1}{2}\eta_{ab}h
\end{equation}
is the trace reversed metric perturbation. Given this expression it
can easily be verified that, for $h_{ab} \equiv 2\eta_{ab}\a$,
\begin{equation}
    -\frac{1}{8\pi}\m{G^{(2)}_{ab}[2\eta\a]}\equiv
    -\frac{1}{2\pi}\m{G^{(2)}_{ab}[\eta\a]} =
    -\frac{3}{4\pi}\m{\a_{,a}\a_{,b}}.
\end{equation}

To conclude, the term $\Sigma^{1(2)}_{ab}$ follows from taking the
second order terms in equation \eqref{s1}. Expressing the covariant
derivatives using the linearized connection
\begin{equation}
    \Gamma^{c(1)}_{ab}:=\frac{1}{2}\eta^{cd}
    \left(\pd_a q_{bd} + \pd_b q_{ad} - \pd_d q_{ab}\right),
\end{equation}
where $q_{ab}:= g^{(1)}_{ab}$ is the first order effective metric
perturbation, gives
\begin{align}\label{ppp}
8\pi\Sigma_{ab}^{1(2)} &:= \left(2\nabla_a \a_{,b} - 2\g \Box\,
\a\right)^{(2)}\nonumber\\
&=-\eta^{cd}\left( \pd_a q_{bd} + \pd_b q_{ad} - \pd_d q_{ab}
\right)\a_{,c}\nonumber\\
&\hspace{.4cm}+ \eta_{ab}\eta^{dc}\eta^{ef}\left( \pd_d q_{cf} +
\pd_c q_{df} - \pd_f q_{dc} \right)\a_{,e}\nonumber\\
&\hspace{.4cm}-2q_{ab}\pd^c\pd_c\a + 2\eta_{ab}q^{dc}\pd_d\a_{,c}.
\end{align}
Using $q_{ab} = \beta_{ab} - 2\a\eta_{ab}$ to eliminate $q_{ab}$,
equation \eqref{ppp} gives a series of terms involving products of
$\beta_{ab}$ and $\a$, whose corresponding averages vanish thanks to
\eqref{su0}. What is left is a sequence of terms, quadratic in $\a$,
which are obtained from \eqref{ppp} upon substitution of $q_{ab}$
with $- 2\a\eta_{ab}$. Their average can be evaluated to yield the
final result
\begin{align}
8\pi\m{\Sigma_{ab}^{1(2)}} = 4\m{\a_{,a}\a_{,b} +
2\eta_{ab}\a^{,c}\a_{,c}}.
\end{align}


\begin{thebibliography}{99}

\bibitem{Unruh1976}
Notes on black-hole evaporation,\\
W. G. Unruh, Phys. Rev. D 14, 870 - 892 (1976)

\bibitem{Ford1975}
Quantum vacuum energy in general relativity,\\
L. H. Ford, Phys. Rev. D 11, 3370 - 3377 (1975)

\bibitem{Ford1993}
Electromagnetic vacuum fluctuations and electron coherence,\\
L. H. Ford, Phys. Rev. D 47, 5571 (1993)

\bibitem{Haisch1994}
Inertia as a zero-point-field Lorentz force,\\
B. Haisch, A. Rueda, and H. E. Puthoff, Phys. Rev. A 49, 678 (1994)

\bibitem{Straumann1999}
The mystery of the cosmic vacuum energy density and the accelerated
expansion of the universe,\\
N. Straumann, Eur. J. Phys. 20 419-427
(1999)

\bibitem{Antoniadis2007}
Cosmological dark energy: prospects for a dynamical theory,\\
I. Antoniadis, P. O. Mazur, E. Mottola, New J. Phys 9, 11 (2007)

\bibitem{Milonni1994}
The quantum vacuum: An Introduction to Quantum Electrodynamics,\\
P. W. Milonni, Academic Press, New York 1994

\bibitem{Rugh02}
The Quantum Vacuum and the Cosmological Constant Problem,\\
S. E. Rugh and H. Zinkernagel, Studies in History and Philosophy of
Modern Physics, 33, 663-705  (2002)

\bibitem{Saunders2002}
Is the zero point energy real?\\
S. Saunders, in: M. Kuhlmann, H. Lyre, A. Wayne (Eds.), Onthological
Aspects of Quantum Field Theory, World Scientific, Singapore  (2002)

\bibitem{Casimir1948}
H. B. G. Casimir, Proc. K. Ned. Akad. Wet. 51, 792 (1948)

\bibitem{Casimir1948b}
The Influence of Retardation on the London-van der Waals Forces,\\
H. B. G. Casimir and D. Polder, Phys. Rev. 73, 360 - 372 (1948)

\bibitem{Mohideen1998}
Precision Measurement of the Casimir Force from 0.1 to 0.9 $\mu$m,\\
U. Mohideen and A. Roy, Phys. Rev. Lett. 81, 4549 - 4552 (1998)

\bibitem{Welton1948}
Some Observable Effects of the Quantum-Mechanical Fluctuations of the Electromagnetic Field,\\
T. A. Welton, Phys. Rev. 74, 1157 - 1167 (1948)

\bibitem{Beiersdorfer2005}
Measurement of the Two-Loop Lamb Shift in Lithiumlike U$^{89+}$,\\
P. Beiersdorfer, H. Chen, D. B. Thorn, and E. Träbert,
Phys. Rev. Lett. 95, 233003 (2005)

\bibitem{Louisell1973}
W. H. Louisell, Quantum Statistical Properties of Radiation, Wiley,
New York (1973)

\bibitem{Gea-Banacloche1988}
Vacuum Fluctuations and Spontaneous Emission in Quantum Optics,\\
J. Gea-Banacloche, M. O. Scully and M. S. Zubairy, Phys. Scr. T21
81-85 (1988)

\bibitem{Wodkiewicz1988}
Stochastic description of vacuum fluctuations,\\
K. W\'{o}dkiewicz, Phys. Rev. A 38, 2932 - 2936 (1988)

\bibitem{Callen1951}
Irreversibility and Generalized Noise,\\
H. B. Callen and T. A. Welton, Phys. Rev. 83, 34 - 40
(1951)

\bibitem{Josephson1962}
Possible new effects in superconductive tunnelling,\\
B. D. Josephson, Phys. Lett. 1, 251 (1962)

\bibitem{Koch1980}
Quantum-Noise Theory for the Resistively Shunted Josephson Junction,\\
R. H. Koch, D. van Harlingen, J. Clarke, Phys. Rev. Lett. 45, 2132
(1980)

\bibitem{Koch1982}
Measurements of quantum noise in resistively shunted Josephson junctions,\\
R. H. Koch, D. van Harlingen, J. Clarke, Phys. Rev. B 26, 74-87
(1982)

\bibitem{Jaffe2005}
The Casimir Effect and the Quantum Vacuum,\\
R. L. Jaffe, Phys. Rev. D 72, 021301(R) (2005)

\bibitem{Doran2006}
What measurable zero point fluctuations can(not) tell us about dark energy,\\
M. Doran, J. Jaeckel 2006, J. Cosmol. Astropart. Phys. JCAP08, 010
(2006)

\bibitem{Fulling2007}
How Does Casimir Energy Fall?\\
S. A. Fulling, K. A. Milton, P. Parashar, A. Romeo, K. V. Shajesh,
and J. Wagner, Phys. Rev. D 76, 025004 (2007)

\bibitem{Beck2005}
Could dark energy be measured in the lab?\\
C. Beck, M. C. Mackey, Phys. Lett. B 605, 295 (2005)

\bibitem{Jetzer2005}
Has dark energy really been discovered in the Lab?\\
P. Jetzer, N. Straumann, Phys. Lett. B 606, 77 (2005)

\bibitem{Jetzer2006}
Josephson junctions and dark energy,\\
P. Jetzer, N. Straumann, Phys. Lett. B 639, 57 (2006)

\bibitem{Mahajan2006}
Casimir effect confronts cosmological constant,\\
G. Mahajan, S. Sarkar, T. Padmanabhan, Phys. Lett. B 641, 6 (2006)

\bibitem{Weinberg1989}
The cosmological constant problem,\\
S. Weinberg, Rev. Mod. Phys. 61, 1-23 (1989)

\bibitem{Mottola1986}
Quantum fluctuation-dissipation theorem for general relativity,\\
E. Mottola, Phys. Rev. D 33, 2136 (1986)

\bibitem{Carroll2001}
The cosmological constant,\\
S. M. Carrol, Living Rev. Rel. 4, 1 (2001)

\bibitem{Zeldovich1967}
Cosmological Constant and Elementary Particles,\\
Y. B. Zel'dovich, JETP letters 6, 316-317 (1967)

\bibitem{Kolb2006}
On cosmic acceleration without dark energy,\\
E. W. Kolb, S. Matarrese and A. Riotto, New J. Phys. 8, 322 (2006)

\bibitem{Li2008}
Is dark energy an effect of averaging?\\
N. Li, M. Seikel, and D. J. Schwarz, Fortschr. Phys. 56, 465- 474
(2008)

\bibitem{CamposVerdaguer1996}
Stochastic semiclassical equations for weakly inhomogeneous cosmologies,\\
A. Campos and E. Verdaguer, Phys. Rev. D 53,
1927 - 1937 (1996)

\bibitem{MartinVerdaguer2000}
Stochastic semiclassical fluctuations in Minkowski spacetime,\\
R. Mart\'in and E. Verdaguer, Phys. Rev. D 61, 124024 (2000)

\bibitem{HuRouraVerdaguer2004}
Induced quantum metric fluctuations and the validity of semiclassical gravity,\\
B. L. Hu, A. Roura, and E. Verdaguer,
Phys. Rev. D 70 (2004)

\bibitem{Hu2004}
Stochastic Gravity: Theory and Applications,\\
B. L. Hu and E. Verdaguer, Living Rev. Relativity 7, (2004), 3.

\bibitem{moffat97}
Stochastic gravity,\\
J. W. Moffat, Phys. Rev. D 56, 6264 - 6277 (1997)

\bibitem{Boyer1975a}
Random electrodynamics: The theory of classical electrodynamics with classical electromagnetic zero-point radiation,\\
T. H. Boyer, Phys. Rev. D 11, 790 - 808 (1975)

\bibitem{Bonifacio2008}
Nonlinear random gravity. II.\\
P. M. Bonifacio, C. H.-T. Wang, R. Bingham and J. T. Mendon\c{c}a,
in preparation.

\bibitem{PowerPercival2000}
Decoherence of Quantum Wave Packets Due to Interaction with Conformal Space-Time Fluctuations,\\
W. L. Power, I. C. Percival, Proc. R. Soc. Lond. A 456, 955 (2000)

\bibitem{Wang2006}
C. H.-T. Wang, R. Bingham and J. T. Mendon\c{c}a,
Classical Quantum Gravity 23, L59-L65 (2006)

\bibitem{Cavalleri1981}
The propagator of stochastic electrodynamics,\\
G. Cavalleri, Phys. Rev. D 23, 363-372 (1981)

\bibitem{Boyer1980}
Thermal effects of acceleration through random classical radiation,\\
T. H. Boyer, Phys. Rev. D 21, 2137 - 2148 (1980)

\bibitem{IbisonHaisch1996}
Quantum and classical statistics of the electromagnetic zero-point field,\\
M. Ibison and B Haisch, Phys. Rev. A 54, 2737 (1996)

\bibitem{Boyer1975b}
General connection between random electrodynamics and quantum electrodynamics for free electromagnetic fields and for dipole oscillator systems,\\
T. H. Boyer, Phys. Rev. D 11, 809 (1975)

\bibitem{York1983}
Dynamical origin of black-hole radiance,\\
J. W. York, Phys. Rev. D 28, 2929 - 2945 (1983)

\bibitem{York2005}
Path integral over black hole fluctuations,\\
J. W. York and B. S. Schmekel, Phys.Rev. D72 (2005) 024022

\bibitem{Isaacson1968}
Gravitational Radiation in the Limit of High Frequency. II. Nonlinear Terms and the Effective Stress Tensor,\\
R. A. Isaacson, Phys. Rev. 166, 1272 (1968)

\bibitem{Flanagan2005}
The basics of gravitational wave theory,\\
E. E. Flanagan, S. A. Hughes, New J. Phys. 7, 204 (2005)

\bibitem{DeWitt1967}
Quantum Theory of Gravity. I. The Canonical Theory,\\
B. S. DeWitt, Phys. Rev. 160, 1113 - 1148 (1967)

\bibitem{Wang2005}
Conformal geometrodynamics: True degrees of freedom in a truly canonical structure,\\
C. H.-T. Wang, Phys. Rev. D 71, 124026 (2005)

\bibitem{ADM1961}
R.Arnowitt, S. Deser, and C.W. Misner Phys. Rev. 121, 1556 (1961)

\bibitem{Hawking&Ellis1973}
The large scale structure of space-time,\\
S.W. Hawking, G.F.R. Ellis, Cambridge University Press (1973)

\bibitem{MTW}
Gravitation,\\
C. W. Misner, K. S. Thorne and J. A. Wheeler, (W. H. Freeman, San
Francisco, 1973)

\bibitem{boomerang}
A measurement of Omega from the North
American test flight of BOOMERANG,\\
A. Melchiorri et al., Astrophys. J., 536, L63 - L66, (2000)

\bibitem{Ia_supernovae}
Observational Evidence from Supernovae for an Accelerating Universe
and a Cosmological Constant,\\
A.G. Riess et al., Astron. J., 116, 1009 - 1038, (1998)

\bibitem{Wang2005b}
Unambiguous spin-gauge formulation of canonical general relativity with conformorphism invariance\\
C. H.-T. Wang, Phys. Rev. D 72, 087501 (2005)

\bibitem{ryder}
Quantum field theory,\\
L. H. Ryder, Cambridge University Press, (1985)

\bibitem{Boyer1969}
Derivation of the blackbody radiation spectrum withut quantum assumptions,\\
T. H. Boyer, Phys. Rev. 182(5), 1374 - 1383 (1969)

\bibitem{Wilczek1999}
Mass Without Mass I: Most of Matter,\\
F. Wilczek, Physics Today 52N11 (1999).

\bibitem{Meissner2007}
Conformal symmetry and the Standard Model,\\
K. A. Meissner, H. Nicolai,
Phys. Lett. B 648, 312–317 (2007)

\bibitem{Higgs1964}
Broken Symmetries and the Masses of Gauge Bosons,\\
P. W. Higgs,
Phys. Rev. Lett. 13, 508 - 509 (1964)

\bibitem{Quigg2007}
Spontaneous symmetry breaking as a basis of particle mass,\\
C. Quigg,
Rep. Prog. Phys. 70, 1019 (2007)

\bibitem{Okada2007}
Higgs Particle: The origin of mass,\\
Y. Okada,
J. Phys. Soc. Japan 76, 111011 (2007)

\bibitem{Wilczek2005}
The origin of mass,\\
F. Wilczek, Mod. Phys. Lett. A21 701-720 (2005).

\bibitem{Gerlich2007}
A Kapitza-Dirac-Talbot-Lau interferometer for highly polarizable molecules,\\
S. Gerlich, L. Hackermüller, K. Hornberger, A. Stibor, H. Ulbricht,
M. Gring, F. Goldfarb, T. Savas, M. M\"uri, M. Mayor, M. Arndt,
Nature Physics 3, 711-715 (2007)

\bibitem{gauge2008}
GAUGE: the GrAnd Unification and Gravity Explorer, G.
Amelino-Camelia, T. Sumner et al, Exp. Astro. DOI:
10.1007/s10686-008-9086-9 (2008)

\end{thebibliography}
\end{document}